\documentclass[fleqn,usenatbib]{mnras}

\usepackage{newtxtext,newtxmath}

\usepackage[T1]{fontenc}

\DeclareRobustCommand{\VAN}[3]{#2}
\let\VANthebibliography\thebibliography
\def\thebibliography{\DeclareRobustCommand{\VAN}[3]{##3}\VANthebibliography}

\usepackage{graphicx}	%
\usepackage{amsmath}	%
\usepackage{multirow}
\usepackage{bigdelim}
\usepackage{physics}
\usepackage{siunitx}
\usepackage{booktabs}
\usepackage[normalem]{ulem}
\usepackage[noabbrev,capitalize]{cleveref}
\crefformat{equation}{Equation~#2#1#3}
\crefrangeformat{equation}{Equations~#3#1#4 to~#5#2#6}
\usepackage[dvipsnames]{xcolor}

\DeclareSIUnit{\MSun}{\ensuremath{M_\odot}} 
\newcommand{\splitcell}[1]{%
  \begingroup
  \renewcommand{\arraystretch}{1}%
  \begin{tabular}{@{}c@{}}#1\end{tabular}%
  \endgroup
}

\title[Lensed gravitational wave localisation]{On the detection and precise localisation of merging black holes events through strong gravitational lensing}

\author[Wempe et al.]{
Ewoud~Wempe,$^{1}$\thanks{E-mail: ewoudwempe@gmail.com} 
L\'eon~V.~E.~Koopmans,$^{1}$ 
A.~Renske~A.~C.~Wierda,$^{2}$ 
Otto~A.~Hannuksela,$^{2,3,4}$ 
\newauthor ~Chris~Van~Den~Broeck$^{2,3}$ \vspace{0.1cm}
\\
$^{1}$Kapteyn Astronomical Institute, University of Groningen, P.O Box 800, 9700 AV Groningen, The Netherlands\\
$^{2}$Institute for Gravitational and Subatomic Physics (GRASP), Department of Physics, Utrecht University, Princetonplein 1, 3584 CC Utrecht, The Netherlands\\
$^{3}$Nikhef – National Institute for Subatomic Physics, Science Park, 1098 XG Amsterdam, The Netherlands\\
$^{4}$Department of Physics, The Chinese University of Hong Kong, Shatin, N.T., Hong Kong 
}

\date{Accepted XXX. Received YYY; in original form ZZZ}

\pubyear{2021}
\newcounter{nameOfYourChoice}

\begin{document}
\label{firstpage}
\pagerange{\pageref{firstpage}--\pageref{lastpage}}
\maketitle
\begin{abstract}
To unlock the full spectrum of astrophysical and cosmological applications of gravitational-wave detections, it is essential to localise the associated black-hole mergers to high precision inside their host galaxies. 
One possible method to achieve this is to compare the properties of multiple detections of gravitationally-lensed binary black-hole merger events with the properties of strong gravitational lens systems located in the joint sky localisation of the gravitational-wave detections.
In this work, we simulate the population of binary black-hole mergers lensed by galaxy-scale lenses and detectable by LIGO-Virgo-Kagra in the coming decade and the population of galaxy-scale strong lenses that will be detected by Euclid.
We use these simulations to investigate the prospects for localising strongly lensed binary black-hole mergers inside the lensed galaxies of ‘Euclid-like’ galaxy-scale strong lenses.
We find that for $20$--$50\,\rm\%$ of strongly lensed gravitational wave events the lens system is detectable with Euclid, if the event falls in its survey footprint.
Of these, we expect to correctly identify the strongly-lensed host galaxy as likely (with posterior probability) host galaxy – based on Bayesian evidence ranking of candidate hosts – for $34.6$--$21.9\,\rm\%$ of quadruply-lensed gravitational-wave events when given an a-priori 1–5 \si{\deg^2} gravitational-wave-only sky localisation.
For triply and doubly lensed gravitational-wave events, this becomes $29.8$--$14.9\,\rm\%$ and $16.4$--$6.6\,\rm\%$ respectively. If successfully identified, however, the localisation can be better than a fraction of the host-galaxy size, i.e. of order milli-arcseconds.
A first detection in the coming decade, however, probably requires dedicated deep and high-resolution follow-ups and continued upgrades in the current and planned gravitational-wave detectors.
\end{abstract}


\section{Introduction}

The LIGO Collaboration and Virgo Collaborations have thus far identified 90 gravitational-wave candidates~\citep{LIGOScientific:2018mvr,LIGOScientific:2020ibl,LIGOScientific:2021djp}. 
Independent analyses of the public data identified additional candidates~\citep{Magee:2019vmb,Nitz:2018imz,Nitz:2020oeq,Nitz:2021uxj,Venumadhav:2019tad,Venumadhav:2019lyq,Zackay:2019btq,Zackay:2019tzo}. 
Further upgrades to LIGO, Virgo, and KAGRA are planned for the upcoming observing runs, and a fifth detector is to be built in India~\citep{Harry:2010zz,Somiya:2011np,LIGOIndia,Aso:2013eba,VIRGO:2014yos,LIGOScientific:2014pky,Akutsu:2020his,Abbott:2020qfu,Saleem:2021iwi}. 
Several new gravitational-wave avenues could become observationally feasible as the detector network improves and expands. One such avenue is the gravitational lensing of gravitational waves. 

In particular, based on the predictions of the expected GW sources and lenses, several independent studies have forecast gravitational-wave lensing observations in the coming years as the detectors are upgraded~\citep{Ng:2017yiu, Li:2018prc, harisIdentifyingStronglyLensed2018, Oguri:2018muv, Smith:2019dis, Xu:2021bfn, wierdaDetectorHorizonForecasting2021}\footnote{However, \citet{yangEventRatePredictions2021} suggests around $0.7$ lensed events per year using a very conservative detection threshold.}. 
Indeed, although the rate of lensing can be subject to uncertainty if the mergers do not directly trace the star-formation rate density~\citep{mukherjeeImpactAstrophysicalBinary2021}, the forecasts are promising. 

As a consequence, rapid progress has been made in strong gravitational-wave lensing analyses \citep{harisIdentifyingStronglyLensed2018,Liu:2020par,Lo:2021nae,janquartFastPreciseMethodology2021,Goyal:2021hxv}, microlensing searches~\citep{Cao:2014oaa,Lai:2018rto,Christian:2018vsi,Singh:2018csp,hannukselaSearchGravitationalLensing2019,Meena:2019ate,Pagano:2020rwj,cheungStellarmassMicrolensingGravitational2021,Kim:2020xkm}, and subthreshold multiple-image searches \citep{liFindingDiamondsRough2021,mcisaacSearchStronglyLensed2020,daiSearchLensedGravitational2020,theligoscientificcollaborationSearchLensingSignatures2021, Goyal:2021hxv}. 
Reliably quantifying and reducing the false alarm probability is actively being worked on too \citep{Caliskan:2022wbh,wierdaDetectorHorizonForecasting2021,janquartFastPreciseMethodology2021}.

These improvements in terms of instruments and methodology open up a unique possibility of a new ``multimessenger''\footnote{Here the term "multimessenger" analysis refers to combining gravitational-wave information (image properties) with electromagnetic information (e.g., lens reconstructions, redshift measurements) in the analysis of a single strong lens galaxy, not the analysis of the compact object binary that emits both gravitational waves and electromagnetic waves.} channel: 

Study of gravitational lens systems using both gravitational-wave and electromagnetic information. 
Such an analysis might be possible if the strong lens galaxy that produces the lensed gravitational wave is localised. 
The localisation in this context could be performed using an "archival" search, by matching the gravitational-wave image properties with electromagnetic lenses: Only the lenses that can produce the observed gravitational-wave properties would be candidate hosts for the binary black hole (BBH). 
Indeed, the multimessenger analysis here refers to the combined analysis of a lens system with multiple messenger signals, as opposed to an analysis using electromagnetic counterparts, such as kilonovae. 
The search can be done with merging compact objects that do not emit detectable electromagnetic counterpart signals, as long as the correct lens system can be discerned among the several strong lenses that exist within the sky localisation of the gravitational wave.

In particular, \citet{Sereno:2011ty,Smith:2017mqu,Robertson:2020mfh,Yu:2020agu,hannukselaLocalizingMergingBlack2020} demonstrated methodologies to combine electromagnetic and gravitational wave measurements to localise the host lens system of merging black holes. 

Indeed, if a gravitational wave from a merging stellar remnant black hole within a galaxy is gravitationally lensed, then the light from within that galaxy is also lensed. 
Because the Earth and thus the detectors will rotate between the arrival time of the different gravitational-wave images, the same BBH is effectively detected several times at different detector orientations, allowing for an effectively "extended" network of detectors and thus significantly improved sky localisation~\cite{Seto:2003iw,hannukselaLocalizingMergingBlack2020}. 
The galaxy would appear as a lensed galaxy within the sky localisation of the wave which will be significantly better localized than a non-lensed gravitational wave~\cite{Seto:2003iw,hannukselaLocalizingMergingBlack2020}. 
Here we vary the sky localization between $\sim 1-5 \, \deg^2$. 
However, an accurate quantification of the sky localization capabilities for variable signal-to-noise ratios, detector networks, and number of strong lensing images, utilizing joint parameter estimation, will be needed in the future, as the sky localization will heavily depend on these assumptions, as has been observed for non-lensed events~\cite{Petrov:2021bqm}.\footnote{For the interested reader, \citet{janquartFastPreciseMethodology2021} shows, a high-SNR example, the sky localization of a quadruply lensed image down to \SI{\sim2}{\deg^2}, using full joint parameter estimation, whereas \citet{hannukselaLocalizingMergingBlack2020} performs a statistical analysis using a simple posterior overlap method, showing that  a large fraction of the events might be constrained to better than 10 $\deg^2$ in the sky, and often to better than 5 $\deg^2$. However, a robust statistical analysis employing joint parameter estimation and duty cycles will be required to better quantify the localization capabilities.}   
Consequently, one can observe this restricted area to find plausible lenses from which the waves originated, resulting in a limited number of candidate lenses (\citet{hannukselaLocalizingMergingBlack2020}, for example, estimate 100-1000 lenses). 
The correct lens system must be able to produce image properties that are consistent with the time delay and magnification ratios detected in the lensed gravitational wave, narrowing down plausible candidates down to (ideally) one. 
A subsequent follow-up analysis of the system could pave the way to a number of scientific applications, similar to the more traditional multimessenger lensing  searches~\citep{Bianconi:2022etr,Ryczanowski:2022int}. 

Crucial for this idea is the ability to detect lenses on the electromagnetic side, and substantial methodological improvements have been made on the lens finding \citep{metcalfStrongGravitationalLens2019}, in particular by incorporating machine learning \citep{petrilloTestingConvolutionalNeural2019,schaeferDeepConvolutionalNeural2018,hartleySupportVectorMachine2017}. %
Observationally, the upcoming Euclid survey \citep{laureijsEuclidDefinitionStudy2011} is expected to be sensitive enough to cover a substantial part of the lensing cross section of galaxies \citep{collettPopulationGalaxyGalaxyStrong2015}.

Unfortunately, it is not clear how practically feasible such a localisation is with near-term instruments like \textit{Euclid} and LIGO-Virgo-Kagra at final sensitivity. 
\citet{hannukselaLocalizingMergingBlack2020} showed that the localisation is possible, in principle, if we assume that we can detect a lensed gravitational-wave event, identify all strong lensing systems within the sky localisation of the event, using three characteristic lens systems as an example and without the full information of the lens and source properties. 
That is, without including a full analysis of the near-term EM instruments' capabilities. 
Here we let go of the latter two assumptions and study how feasible the localisation is in practice with \textit{Euclid} and LIGO--Virgo--Kagra at final sensitivity, given that we have detected a strongly lensed gravitational-wave event from a galaxy lens.
In addition, we extend the Bayesian analysis framework to account for a more realistic astrophysical scenario.

We restrict ourselves to galaxy-galaxy lensing, because the lens morphology is simpler to describe.

In particular, we perform a population study and a full multidimensional Bayesian modelling of time-delays and magnifications, including also our prior expectation for the galaxy's stellar mass content. 
In addition, we perform the full electromagnetic modelling and consider the degeneracies that arise from this. 
Furthermore, we analyse a substantial number of simulated lenses inside the footprint to estimate performance. 
We focus on using \textit{Euclid}-like imaging, and we will in practice thus restrict to events in \textit{Euclid} fields, although with a ground-based telescope like Subaru/HSC\footnote{\url{https://www.naoj.org/Projects/HSC/}}or Rubin/LSST\footnote{\url{https://rubinobservatory.org/}}, there are still good options if the source is not in the \textit{Euclid} footprint. 
Additionally, the \SI{\sim40}{\percent} footprint of \textit{Euclid} might be supplemented with the \SI{17500}{\deg^2} footprint of the also upcoming Chinese Space Station Telescope \citep{gongCosmologyChineseSpace2019}.

To test the framework, in \cref{sec:simulations}, realistic simulations of lensed gravitational waves and lens systems are set up. The results and realisations of these simulations are shown in \cref{sec:samplegen}. 
In \cref{sec:gwloc}, we work out a procedure for the identification and localisation of lensed GWs, and apply it on a set of simulated lenses to test what accuracies we might achieve in practice. 
In \cref{sec:discussion}, we summarise and discuss our findings.

Throughout this work, we assume the Planck18 cosmology \citep{planckcollaborationPlanck2018Results2020}.

\section{Simulations}
\label{sec:simulations}
In this section, realistic simulations are created of the gravitational lenses, the host galaxies, and the lensed gravitational waves. This is mostly analogous to \citet{collettPopulationGalaxyGalaxyStrong2015}, but includes GWs. A summary of the simulation parameters to supplement the text is in \cref{tab:summarysims}. We first generate the lens (\cref{sec:lensmass,sec:lenslight}) and source populations (\cref{sec:sourcepop}), then the gravitational wave populations (\cref{sec:gwpop}). The gravitational waves are placed in source galaxies as described in \cref{sec:gwpos}. In \cref{sec:imgreq}, the detectability criteria of lens systems by imaging surveys are defined. Finally, in \cref{sec:lenssysprob}, the lensing probability is discussed. 
\subsection{Lens mass model}
\label{sec:lensmass}
\begin{figure}
    \centering
    \includegraphics[width=\linewidth]{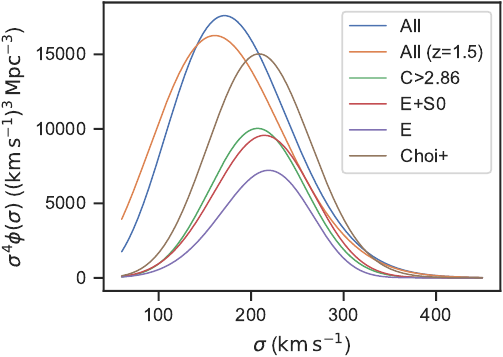}
    \caption{Various weighted measured velocity dispersion functions from \citet{bernardiGalaxyLuminositiesStellar2010}, and \citet{choiInternalCollectiveProperties2007}. The lensing cross section scales roughly like $\propto\sigma^4$. The lines show the distributions for all galaxies (blue), all galaxies with the redshift correction \cref{eq:veldispfunc} (orange), galaxies with a concentration threshold (green), a sample targeting S0's and E's (orange), a (smaller) morphologically selected sample of E's + S0's (red), just ellipticals (purple) and a selection of early-type galaxies (brown). Except one (brown) all distributions are taken from \citet{bernardiGalaxyLuminositiesStellar2010}. In this paper, we utilise the All (\emph{All galaxies} in \cref{tab:rates})) and Choi+ (\emph{Early-type galaxies}).) velocity dispersion functions.}
    \label{fig:vdfs}
\end{figure}

The lens population mostly follows \citet{oguriEffectGravitationalLensing2018} and \citet{collettPopulationGalaxyGalaxyStrong2015}.
For the lens galaxy velocity dispersion $\sigma_l$ and lens redshift $z_l$ we follow the procedure of \citet{oguriEffectGravitationalLensing2018}, who takes a local velocity dispersion function measurement $\phi_\text{loc}$ from \citet{bernardiGalaxyLuminositiesStellar2010} and combines it with the redshift evolution of the velocity dispersion function of Illustris \citep{genelIntroducingIllustrisProject2014,vogelsbergerPropertiesGalaxiesReproduced2014,genelIntroducingIllustrisProject2014} galaxies $\phi_\text{hyd}$ as characterised by \citet{torreyAnalysisEvolvingComoving2015}
\begin{align}
\phi(\sigma_l, z_l) = \phi_\text{loc}(\sigma_l)\frac{\phi_\text{hyd}(\sigma_l,z_l)}{\phi_\text{hyd}(\sigma_l,0)}.\label{eq:vdf}
\end{align}
We verify that the implementation matches Figure 1 of \citet{oguriEffectGravitationalLensing2018}.
Including the comoving volume element gives the joint probability density function $\pi_l$:
\begin{align}
\pi_l(\sigma_l, z_l) = \phi(\sigma_l,z_l) \dv{V(z_l)}{z_l},\label{eq:veldispfunc}\\
\dv{V(z_l)}{z_l} = 4\uppi \frac{c}{H_0} \frac{(1+z_l)^2D_l^2}{E(z_l)}.
\end{align}
As limits, we use $60<\sigma_l/(\si{km/s})<600$ and $0<z_l<3$.\footnote{Note that the expected velocity dispersions and lens redshifts span a much narrower region (e.g., \cref{fig:vdfs}), and thus the limits are justified.} $\pi_l$ is then normalised over these limits.
The choice of the velocity dispersion function can significantly impact the expected populations, as illustrated in \cref{fig:vdfs}, where various weighted velocity dispersion functions are shown. These velocity dispersion functions are derived from stellar velocity dispersions, but we use them for the mass model too, which is consistent with \citet{boltonSloanLensACS2008}.

The mass model is an elliptical power law based on the formalism of \citet{tessoreEllipticalPowerLaw2015}, see also \citet{barkanaFastCalculationFamily1998}, with Einstein radius \begin{equation}\theta_E = \frac{4\uppi \sigma_l^2}{c^2}\frac{D_\text{LS}}{D_\text{OS}},\end{equation} where $D_\text{LS}$ and $D_\text{OS}$ are angular diameter distances between Lens and Source and between Observer and Source respectively.
The centroid of the lens is slightly perturbed such that it is not at a fixed location with respect to the pixel boundaries.
    We follow \citet{collettPopulationGalaxyGalaxyStrong2015} and assume a Rayleigh distribution for their ellipticities 
\begin{equation}
    p(q) = \frac{1-q}{s}\exp(-(1-q)^2/2s^2)\label{eq:ellipticity}
\end{equation}
truncated at $q=0.2$, with a scale $s$ of\footnote{There is a typo in the corresponding equation \citet{collettPopulationGalaxyGalaxyStrong2015} (T.~Collett 2021, private communication), so we use the correct equation from their code LensPop, which, with increasing velocity dispersion, has a decreasing instead of an increasing ellipticity.}
\begin{equation}s=0.378-\num{5.72e-4}\sigma/(\si{km.s^{-1}}).\label{eq:ellipticityscale}\end{equation}
This is equivalent to defining their complex ellipticity components $e_1$ and $e_2$ to have Gaussian distributions.
Although on average the light and mass ellipticity vectors are aligned for SLACS lenses \citep{shuSloanLensACS2015}, there is still some scatter, so we add this scatter to the drawn light ellipticities:
\begin{align}
q_\text{mass}/q_\text{light}\sim\mathcal{N}(1,\sigma_f^2)\\
\theta_\text{mass}\sim\mathcal{N}(\theta_\text{light}, \sigma_\theta^2)
\end{align}
The power-law index is drawn from \begin{equation}\gamma_m \sim \mathcal{N}(2,\sigma_\gamma^2),\end{equation} a singular isothermal ellipsoid would have $\gamma_m=2$, a point mass has $\gamma_m=3$ and a mass sheet has $\gamma_m=1$.
We use $\sigma_\gamma=0.2$.
Although we assume this on the astrophysical distribution of $\gamma_m$ rather than the observed distribution (i.e. the distribution after selection effects, magnification bias, etc.), we found that after drawing our sample of lenses that match our detectability criteria, the distribution of $\gamma_m$ in our simulations matches well with the observed distribution of \citet{augerSloanLensACS2010}.
We also include an external shear, with magnitudes from a Rayleigh distribution with scale $s=0.05$, and random lens galaxy position angles as in \citet{collettPopulationGalaxyGalaxyStrong2015}.
We note that this external shear should not be interpreted as mostly originating from external matter \citep{etheringtonExternalShearIsNotShear}. Instead, it should be considered a first-order correction or first-order deviation of the lens model. For instance, it can absorb a galaxy being disky or boxy.

\subsection{Lens galaxies}
\begin{table*}
    \centering
    \caption{Summary of the model. See the full text and referred sections for more complete explanations.}
\begin{tabular}{
  l
  l
  l
  l
}
\toprule
{Model part} & {Parameter} & {Functional Form} & {Values and comments} \\
\midrule
Lens mass model & Velocity dispersion $\sigma$ \rdelim\}{2}{0mm} & \multirow{2}{*}{\cref{eq:veldispfunc}} & \multirow{2}{*}{All-galaxy velocity dispersion function, see text}\\
\cref{sec:lensmass}       & Redshift $z_l$ &  & \\
             & Mass axis ratio $q_\text{mass}$ & $q_\text{mass}/q_\text{light}\sim\mathcal{N}(1,\sigma_f^2)$& $\sigma_f=0.2$\\
             & Mass axis angle $\theta_\text{mass}$ & $\theta_\text{mass}\sim\mathcal{N}(\theta_\text{light}, \sigma_\theta^2)$ &  $\sigma_\theta=\SI{34}{\degree}$\\
             & Power-law index $\gamma$ & $\mathcal{N}(\mu_\gamma,\sigma_\gamma^2)$ & $\mu_\gamma=2, \sigma_\gamma=0.2$\\
             & Shear $\gamma_s$ & Rayleigh & Scale $s=0.05$\\ %
             & Shear angle $\theta_s$ & $\mathcal{U}(0,2\pi)$ & \\
             & Lens position $\vb*x_l$ & $\mathcal{U}(\vb*x_l)$ & Varied within the central pixel\\
\midrule
Lens light model & Absolute magnitude $M_r$ \rdelim\}{2}{0mm}& \multirow{2}{*}{\cref{eq:magnitude}} & \multirow{2}{*}{Fundamental plane, see text}\\
\cref{sec:lenslight} & S\'ersic radius $r_l$ & & \\
&$k$-correction $k$ & $k\sim\mathcal{N}\qty(\mu_k(M_r, z_l), \sigma_k(M_r, z_l))$ & Distribution following Millennium galaxies (see text)\\
& S\'ersic index $n_l$ & $n_l=4$ & Fixed\\
& Light axis ratio $q_\text{light}$ & Rayleigh, \cref{eq:ellipticity} & Scale $s$ from SDSS fit by \citet{collettPopulationGalaxyGalaxyStrong2015} (\cref{eq:ellipticityscale})\\
& Light axis angle $\theta_\text{light}$ & $\mathcal{U}(0,2\pi)$ & \\
\midrule
Source model & Stellar mass $\log_{10}{M_*}$\hfill\rdelim\}{7}{0mm} & \multirow{7}{*}{Density estimate} & \multirow{7}{*}{JAGUAR catalogue \citep{williamsJWSTExtragalacticMock2018}} \\
\cref{sec:sourcepop}& Star formation rate $\text{SFR}$ & & \\
& Redshift $z_s$ & & \\
& VIS magnitude $m_\text{VIS}$ & & \\
& S\'ersic radius $r_s$ & & \\
& Axis ratio $q_s$ & & \\
& S\'ersic index $n_s$ & & \\
& Axis angle $\theta_s$ & $\mathcal{U}(0,2\pi)$ & \\
& Source position  $\vb*{x}_s$ & \cref{eq:priorsource} & Uniformly distributed\\
\midrule
GW model & Masses $m_1, m_2$ & \textit{Power Law + Peak} & O1-O3a fit from \citet{abbottPopulationPropertiesCompact2021}\\
\cref{sec:gwpop}& Hour angle HA & $\text{HA}\sim\mathcal{U}(0,2\pi)$  \hfill \rdelim\}{4}{0mm} & \multirow{4}{*}{Natural priors}\\
& Declination $\delta$ & $\delta=\arcsin(2u-1),u\sim\mathcal{U}(0,1)$ \\
& Polarisation angle $\psi$ & $\psi\sim\mathcal{U}(0,2\pi)$           \\
& Inclination $\iota$ & $\iota=\arccos(1-2u),u\sim\mathcal{U}(0,1)$   \\
& Waveform model & IMRPhenomD & from \texttt{pycbc} \\
& Noise PSD & aLIGOZeroDetHighPower & aLIGO design sensitivity, from \texttt{pycbc} \\
& Detectors & Hanford, Livingston and Virgo & \\
\cref{sec:gwpos}& Position $\vb*{x}_\text{BBH}$ & \cref{eq:gwposgivensource} & Following light distribution\\
\bottomrule
\end{tabular}
\label{tab:summarysims}
\end{table*}

\label{sec:lenslight}
To each lens, a lens galaxy is coupled, with the centroid of the lens mass profile matching the centroid of the lens galaxy light profile.
To describe the light distributions of the lens galaxies, we assume that they are early-type galaxies that follow the rest-frame fundamental plane. Specifically, we follow the approach of \citet{goldsteinRatesPropertiesSupernovae2019}. This assumption is suitable mainly for ellipticals (and to a lesser degree for lenticulars), but only a small amount of the lensing cross section is due to spirals \citep{sygnetSearchEdgeonGalaxy2010,feronSearchDiskGalaxyLenses2009}.
The lens-galaxy light brightness profile parameters are distributed following the rest-frame Fundamental Plane, conditional on the sampled velocity dispersion from the velocity dispersion function. 
Unlike \citet{goldsteinRatesPropertiesSupernovae2019}, we choose to work with the FP in the $r$-band instead of the $i$-band, and additionally, we use the FP in the lens galaxy rest-frame. The FP is characterised as a multivariate Gaussian described by the surface brightness, $\mu$, the log of the effective radius, $\log_{10}{R_l}$, and the log of the velocity dispersion, $\log_{10}\sigma_l$, as obtained by \citet{bernardiEarlyTypeGalaxiesSloan2003}.
Because the velocity dispersion was already sampled using \cref{eq:veldispfunc}, $\mu$ and $R_l$ are drawn from this Gaussian conditional on the sampled velocity dispersion.\footnote{There is a typo in \citet{goldsteinRatesPropertiesSupernovae2019}. Specifically, the mean conditional on the velocity dispersion should be  $\bmqty{\mu_{*,c}\\R_*} + \frac{V-V_*}{\sigma_V}\bmqty{\sigma_\mu\rho_{V\mu}\\\sigma_R\rho_{RV}}$.}
The resulting surface brightness and half-light radii are converted to an absolute magnitude as follows,
\begin{equation}
    M_r = \mu - 5\log_{10}(R_e/ (h_{70}(\SI{10}{pc})(\SI{1}{arcsec})))-2.5\log_{10}(2\uppi),\label{eq:magnitude}
\end{equation}
where $D_\text{lum}$ is the luminosity distance. These can be derived by rewriting Equation (24) of \citet{goldsteinRatesPropertiesSupernovae2019}.

To generate images, we need the apparent magnitude $m$ of the lens galaxies, which depends on the galaxy's absolute magnitude $M$, its distance $d$ and a $k$-correction: $m = M + 5 \log_{10} (d/\SI{10}{pc}) + k$.
The $k$-correction is required because of the difference in brightness due to the shifting of the spectrum by the galaxy's redshift, while the filter remains constant.
Since the behaviour depends on the spectra of galaxies, we derive them from galaxies in a Millenium simulation lightcone of \citet{henriquesGalaxyFormationPlanck2015}.
We assume that they only depend on the redshift and absolute r-band magnitude.
Specifically, a probability distribution was made of the mean and standard deviation of the k-corrections in a certain $M_r$, $z$ bin, giving $\mu_k(M_r, z)$ and $\sigma_k(M_r,z)$.
To smoothen the distribution, a third-order polynomial fit was made on this histogram.
The $k$-corrections are then drawn from a Gaussian given the calculated mean and standard deviation.
Modelling the $k$-correction as a Gaussian with these simulation-derived standard deviations as spread is necessary to account for the fact that observationally, the $k$-correction is uncertain, especially at higher redshifts (although the bulk of our lenses has redshifts below 1, see \cref{fig:lenscorner}).
We do not include the $k$-correction required for the mismatch between the $r$-band and the VIS-band. However, we note that our results are robust against the exact treatment of the light distributions of lens galaxies, because in the lens modelling, lens galaxies are subtracted anyway.

In the simulations, we assume that the actual light distribution is an elliptical de Vaucouleurs profile (S\'ersic profile with $n_l=4$). This assumption is reasonable for most lenses, which are massive early-type galaxies. In the lens modelling, however, the S\'ersic index is a free parameter, with a broad uniform prior. We note that this, and other simplifying assumptions on the lens galaxies, only enter in the identification in the process of lens-galaxy subtraction and modelling, so the effects are considered second order.

\subsection{Source galaxies}
\label{sec:sourcepop}
\begin{figure*}
    \centering
    \includegraphics{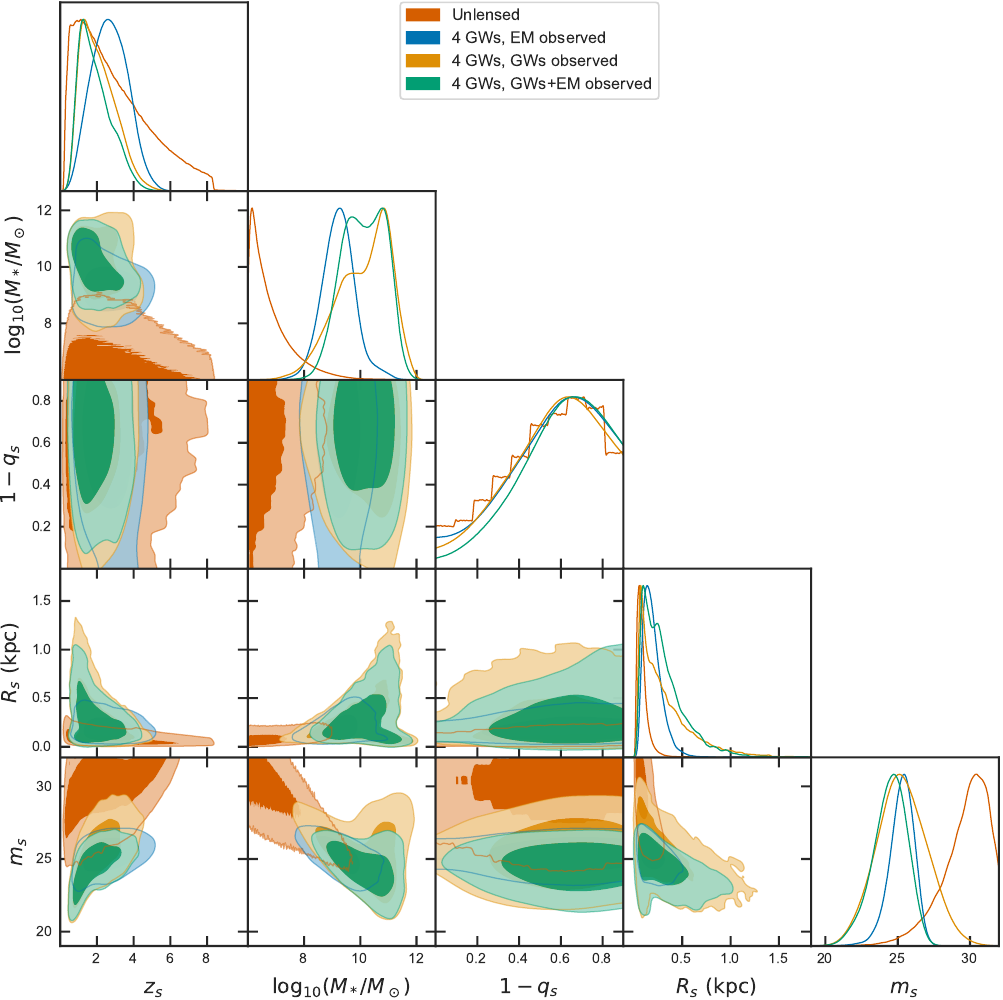}
    \caption{Source galaxy parameter distributions of the simulated sample of lens systems that could host a quadruply-lensed GW. Colors are like in \cref{fig:lenscorner}, with the addition of the red-brown distribution, which is the distribution of the intrinsic host galaxy population (the Jaguar catalogue \citep{williamsJWSTExtragalacticMock2018}). Shown are the source redshift ($z_s$), the galaxy's stellar mass ($\log_{10}(M_*/M_\odot$), the galaxy's ellipticity ($1-q_s$), the effective radius $R_s$, and the VIS-magnitude ($m_s$). The bin-like structure for the ellipticity is an artifact of the generation of the Jaguar catalogue \citep{williamsJWSTExtragalacticMock2018}, generated as described in \cref{sec:physrealisations}.}
    \label{fig:sourcecorner}
\end{figure*}

Considering that the magnifications due to lensing can make sources below the usual magnitude cutoff visible, the regular \textit{Euclid} simulations do not go deep enough for our purposes. 
Thus, we use the simulated, but observationally motivated JAGUAR catalogue \citep{williamsJWSTExtragalacticMock2018}, constructed as forecast for the planned deep JADES survey on the James Webb telescope. It is a complete catalogue of galaxies between $0.2<z<15$, and a stellar mass cutoff of $M_*>10^6\;\si{\MSun}$ (at high redshift this is converted into an equivalent UV-magnitude cutoff), such that there is no direct selection on the source magnitude. This catalogue includes stellar masses, S\'ersic profile parameters and spectra. We integrated the spectra with the VIS filter to get VIS magnitudes. Galaxies with VIS-magnitudes above 32 are filtered out (this is less than \SI{1}{\percent} of the simulation's stellar mass). We use a denoising density estimation tool \citep{bigdeliLearningGenerativeModels2020} to construct a smooth probability distribution function from this catalogue. Specifically, the sampled parameters are  redshifts ($z_s$), stellar masses ($M_*$) or star-formation rates (SFR), VIS magnitudes ($m_s$), effective radii $R_s$, axis ratios $q_s$ and S\'ersic indices $n_s$. The distributions of these parameters are shown in \cref{fig:sourcecorner}. The ellipticity angle is isotropically distributed. 

\subsection{Gravitational wave population}
\label{sec:gwpop}
To generate realistic lensed GWs, we must couple a GW event to a source galaxy. The redshift is set equal to the source galaxy redshift. To account for the fact that the probability of emitting a GW varies per galaxy, it depends on the galaxy parameters, we include a GW emission probability, which is described in \cref{sec:lenssysprob}.
We note that although this is a different way of sampling the gravitational wave redshifts, we find that the redshift distributions are very similar to what was found by \citet{wierdaDetectorHorizonForecasting2021}, who directly sample the gravitational wave redshifts using distributions that come from the redshift evolution in stellar population models, instead of linking them to galaxies.
For the other observed gravitational wave parameters, we follow the procedure of \citet{wierdaDetectorHorizonForecasting2021}, who use the observed LIGO-Virgo-Kagra O1-O3a distributions. We adopt a stellar remnant black mass function based on the \textit{Power Law + Peak} model fitted to the LIGO/Virgo detections across O1-O3a by \citet{abbottPopulationPropertiesCompact2021}. Using these masses, the resulting signal-to-noise ratio (SNR) is calculated for an ideally oriented source, using the \texttt{pycbc} package \citep{nitzGwastroPycbcPyCBC2021}, for A+ sensitivities for LIGO, and design sensitivities for Virgo and KAGRA. 
The antenna pattern, which is a function of the position on the sky $(\alpha,\delta)$, the event time, and the inclination (the angle between the orbital angular momentum and the line of sight), includes 4 GW detectors (Hanford--Livingston--Virgo--KAGRA). This gives the intrinsic SNR of a GW (i.e., before any lensing is applied). The lensed SNRs of each image $i$ is then the original signal-to-noise ratio times the square root of the magnification ($\rho_i=\rho_{\rm intrinsic}\sqrt{\mu}$)~\cite[e.g.][]{Ng:2017yiu}. 

Here we assume 100\% duty cycle for simplicity. 
However, in reality, the duty factor will alter the absolute rate of detections and thus the rate of gravitational-wave lensing detections. 
The detection efficiency, likewise, will depend on the number of detectors that are online, as a more extensive detector network will allow for better sky localization and thus a greater lensing detection efficiency.  
In O3, the duty factor for at least one detector being online was 97\%; for any two detectors being online at the same
time it was 82\%; and for all three detectors together it was
45\%. Further details regarding instrument performance and
data quality for O3a are available in~\cite{LIGOScientific:2020ibl}. 
Time not observing is spent
either acquiring lock, unlocked and undergoing maintenance, unlocked due to unfavorable environmental conditions (earthquakes, wind, storms), or locked and in a state
of commissioning, where improvements are made to the
detectors. If a subset of the gravitational-wave detectors is online, the signal-to-noise ratio will be lower, which will somewhat degrade the rate of detectable lensed events. 
However, the duty cycle factor is being constantly improved (see e.g.,~\cite{aLIGO:2020wna} for improvements from 2015 to 2022), and we furthermore conservatively do not include the LIGO India detector, a fifth detector, that is being built in India.\footnote{For example, presuming a final duty factor of 80\% and a 5-detector network, four or more detectors would be online 73\% of the time. }

\subsection{Binary black hole source position} 
\label{sec:gwpos}
An astrophysical assumption that we make is that within the galaxy, the BBH source position prior follows the observed stellar VIS-band light distribution. This is a simplified model of reality: ideally, one would consider the rest-frame light distribution, in possibly some other band, depending on the exact (currently mostly unknown) astrophysical nature of the BBH population. For instance, if the BBH rates are linked to more massive stars, the blue light distribution would represent the position dependence of the BBH rate.

Given this assumption, the prior probability of the BBH position inside the galaxy, $\vb*{x}_\text{BBH}$, is proportional to the source light profile. In a model where the gravitational wave population follows the stellar mass distribution, this is equivalent to assuming a constant mass-to-light ratio inside the galaxy. 
We furthermore sample conditional on a minimum number of images for the GW. For instance, when simulating systems with $\num{\ge4}$ images, a priori, the gravitational wave is constrained to be inside the diamond four-image caustic.

The first way of getting realisations that satisfy these two criteria, namely having the BBH source position being distributed like the light distribution, and constraining the number of images, is starting from source galaxy position $\vb*x_s$ and sampling the BBH source position conditional on this source position. The priors are
\begin{align}
    &\pi(\vb*x_s) = \mathcal{U}(\vb*x_s) = 1/(4\uppi) \qq{(flat prior)}\label{eq:priorsource}\\
    &\pi(\vb*x_\text{BBH}|\vb*x_s, \vb*\theta_l, \vb*\theta_s, \text{GW in caustic}) =\nonumber\\&\hspace{2cm}
    \frac{\mathrm{S\acute{e}rsic}(\vb*x_s|\vb*\theta_s, \vb*x_\text{BBH})\mathcal{U}_\diamondsuit(\vb*x_\text{BBH})}{\int\dd{\vb*x_\text{BBH}'}\mathrm{S\acute{e}rsic}(\vb*x_s|\vb*\theta_s, \vb*x_\text{BBH}')\mathcal{U}_\diamondsuit(\vb*x_\text{BBH}')}.\label{eq:gwposgivensource}
\end{align}
The denominator, which is the probability that a BBH, whose source position follows the light distribution of a particular source galaxy, is inside the diamond caustic, must also be taken into account:
\begin{align}
    &p(\text{GW in caustic}|\vb*x_s, \vb*\theta_l, \vb*\theta_s) =\nonumber\\&\hspace{2cm}\int\dd{\vb*x_\text{BBH}'}\mathrm{S\acute{e}rsic}(\vb*x_s|\vb*\theta_s, \vb*x_\text{BBH}')\mathcal{U}_\diamondsuit(\vb*x_\text{BBH}') \label{eq:pgwincaustic}
\end{align}

The second way is by first sampling the BBH source position uniformly within the diamond caustic ($\mathcal{U}_\diamondsuit$, which depends on the lens parameters $\vb*\theta_l$ and source parameters $\vb*\theta_s$), and only afterward sampling the galaxy source position. Effectively, the BBH source position and source galaxy source position are switched; this is possible because we have a flat prior on the source position. The priors are: 
\begin{align}
    \pi(\vb*x_\text{BBH}|\vb*\theta_l,\vb*\theta_s) &= \mathcal{U}_\diamondsuit(\vb*x_\text{BBH}|\vb*\theta_l, \vb*\theta_s)\label{eq:gwprior}\\
    \pi(\vb*x_{s}|\vb*x_\text{BBH}, \vb*\theta_s, \vb*\theta_s) &= \frac{\mathrm{S\acute{e}rsic}(\vb*x_\text{BBH}|\vb*\theta_s,\vb*x_s)\pi(\vb*{x}_s)}{\int\dd{\vb*x_s'}{\mathrm{S\acute{e}rsic}(\vb*x_\text{BBH}|\vb*x_s',\vb*\theta_s)\pi(\vb*{x}_s')}}\\
    &=\mathrm{S\acute{e}rsic}(\vb*x_\text{BBH}|\vb*\theta_s,\vb*x_s)\pi(\vb*{x}_s)\label{eq:priorgwpos},
    \end{align}
where this last step was made by assuming a flat prior on the source position ($\pi(\vb*x_s)=1/(4\uppi)$), and defining the S\'ersic function to be normalised.

With some work, \cref{eq:gwprior,eq:priorgwpos} can be sampled from with simple inverse transform sampling. For forward simulations, the second method (using \cref{eq:priorgwpos}) is most convenient, but for doing identification, we require the first method. In practice, the caustics are calculated analytically using the equations from \citet{tessoreEllipticalPowerLaw2015}, as detailed in \cref{sec:caustics}. For the lens equation solving (necessary for calculating magnifications and time delays), a semi-analytical recipe was used too, outlined in \cref{sec:lensequationsolving}. 

\subsection{Gravitational wave detectability and lens identifiability}
\label{sec:imgreq}
\label{sec:gwreq}
Although in practice GW detections are classified as detection using the false alarm rate, we opt for a hard SNR threshold, which can be used as a reasonable proxy for a detection. %
Targeted lensed searches, when one or more super-threshold image is available, may allow one to uncover so-called sub-threshold triggers below the usual noise threshold by reducing the background noise and glitch contribution~\citep{liFindingDiamondsRough2021,mcisaacSearchStronglyLensed2020}. 
For the choice of threshold, we repeat the argument of \citet{wierdaDetectorHorizonForecasting2021}: usually, a trigger requires $\text{SNR}>8$ \citep{abbottProspectsObservingLocalizing2020}, but work on sub-threshold triggers by \citet{liFindingDiamondsRough2021} found a detector horizon distance (the furthest distance a detector network will be able to detect an event) increase of \SI{\sim15}{\percent}. To take this into account, we choose as threshold $\text{SNR}=7$ in our work (note that the $\text{SNR}\propto d_L^{-1}$). 
Planned improvements to the sub-threshold search pipelines are expected to push the sensitivity down further. 

To determine whether a lens system can be identified as such in \textit{Euclid} images, we opt for hard thresholds on some the following parameters (see \cref{app:observability} for further justifications on the numbers):
\begin{enumerate}
    \item The solid angle over which the flux from the background galaxy is detected at $\num{\ge1.5}\sigma$ per pixel exceeds \SI{0.70}{\arcsec^2}. This area is a measure of the information content that an image has -- and is a reasonable measure of its difficulty to be detected by lens finders \citep{nagamdenselens}. Specifically, it corresponds to an area of 30 resolution elements (of $\pi \qty(\text{FWHM}/{2})^2$).
    \item The system is sufficiently resolved, this requires an Einstein radius above \SI{0.33}{\arcsec}. 
    \item The background galaxy brightness distribution varies sufficiently across the lens: if the lens is small compared to the angular scale at which the source galaxy light profile varies, then the deflection of the light rays due to the lens does lead to a noticeable distortion of the source galaxy, and therefore it is undetectable. We found that the parameter that describes this scale best is the semi-minor axis length, because this is the shortest length scale across which the galaxy varies, and we require $r_m<0.56\theta_E$ (see \cref{app:observability}). 
\end{enumerate}
The probability of detection is therefore a sharp cut:
\begin{equation}
    P_\text{EMdet}(\vb*\theta_l,\vb*\theta_s)=\begin{cases}1 &\qq{if lens system identifiable}\\0&\qq{else}\\\end{cases}
\end{equation}
\begin{figure}
    \centering
    \includegraphics[width=\linewidth]{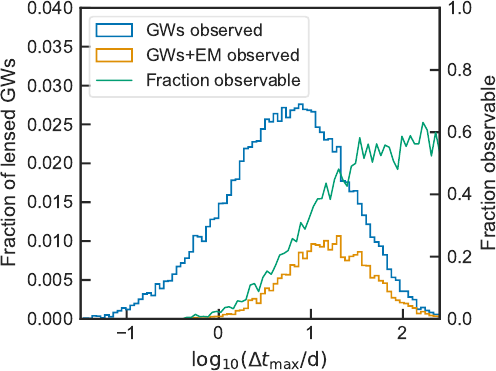}
    \caption{The time delay distribution (we take here the time between the first and last event) for 4 detected GWs with $M_*$ as GW rate. The green line shows the fraction of lenses that are also observable in the EM in each bin.}
    \label{fig:obs_time}
\end{figure}

\subsection{Lensing and detection probabilities}
\label{sec:lenssysprob}
Not all galaxies are equally likely to host BBHs, nor are all galaxies equally likely to lens the BBH.\footnote{For example, a galaxy with more stellar matter is more likely to host the binary black hole, and a more massive galaxy is more likely to strongly lens the binary black hole than a lighter one.} 
We combine all the previous ingredients to finally fold in our analysis the probability that a given system hosts a strongly lensed BBH. 
\begin{enumerate}
    \item The first factor is a geometrical term and concerns the lensing cross section, i.e. the area within the relevant caustic in the source plane $\sigma_c(\vb*\theta_l,\vb*\theta_s)$. The probability density\footnote{In this manuscript, we denote probability densities with $p$ and probabilities (that are between 0 and 1) with (capital) $P$.} of a random GW event in the sky to occur inside that caustic is
    \begin{equation}
        p_\text{cross sec}(\vb*\theta_l,\vb*\theta_s) = \frac{\sigma_c(\vb*\theta_l,\vb*\theta_s)}{4\uppi}.
    \end{equation}
    The lens and source parameters $\vb*\theta_l$ and $\vb*\theta_s$ follow the PDFs as earlier discussed.
    \item Additionally there is the probability that a particular source galaxy with parameters $\vb*\theta_s$ emits a gravitational wave. What this exactly looks like is quite uncertain. Predictions based on stellar population modelling vary \citep{artaleHostGalaxiesMerging2019,santoliquidoCosmicMergerRate2021} and observationally the redshift evolution is most likely bracketed between the star formation rate and the stellar mass density \citep{abbottPopulationPropertiesCompact2021}. We, therefore, consider two models\footnote{The SFRs and stellar masses that are inputs to these models are obtained as part of our sampling of source galaxies, see \cref{sec:sourcepop}.}: one where the GW emission probability is proportional to the stellar mass of the source galaxy, 
    \begin{equation}
        p_{\text{emitted GW}, M_*}(\vb*\theta_s) = A_{M_*} M_*,
    \end{equation}
    and one where the probability is proportional to the galaxy's star formation rate (in the last \SI{100}{Myr}), 
    \begin{equation}
        p_\text{emitted GW, \text{SFR}}(\vb*\theta_s) = A_\text{SFR} \text{SFR},
    \end{equation}
    where $A_{M_*}$ and $A_\text{SFR}$ are constants of proportionality (e.g. the number of GWs per year per solar mass).

A rough estimation of the number of detections can be obtained by dividing the BBH merger rate as measured by \citet{abbottPopulationPropertiesCompact2021} by the stellar mass density (or SFR) at $z=0$ as assumed in \citet{williamsJWSTExtragalacticMock2018}, which gives 
\begin{align}
    &A_{M_*} = \frac{\SI[parse-numbers=false]{23.9\substack{+14.3\\-8.6}}{Gpc^{-3}.yr^{-1}}}{\SI{\sim1.8e8}{\MSun.Mpc^{-3}}} \sim \SI{1.3e-16}{yr^{-1}.\MSun^{-1}}\\
    &A_\text{SFR} = \frac{\SI[parse-numbers=false]{23.9\substack{+14.3\\-8.6}}{Gpc^{-3}.yr^{-1}}}{\SI{\sim1.5e-2}{\MSun.yr^{-1}.Mpc^{-3}}} \sim \SI{1.6e-6}{yr^{-1}.(\MSun.yr^{-1})^{-1}}
\end{align}
    \item To consider only detectable lenses there is a probability that we observe the system with an electromagnetic (EM) imaging survey, $p_\text{EMdet}$. The detectability criterion of \cref{sec:imgreq} is applied. 
    \item To consider only detectable gravitational waves, there is the detection probability $p_\text{GWdet}$, as explained in \cref{sec:gwreq}. %
\end{enumerate}

\section{Sample and data generation}
\label{sec:samplegen}
\begin{figure*}
    \centering
    \includegraphics{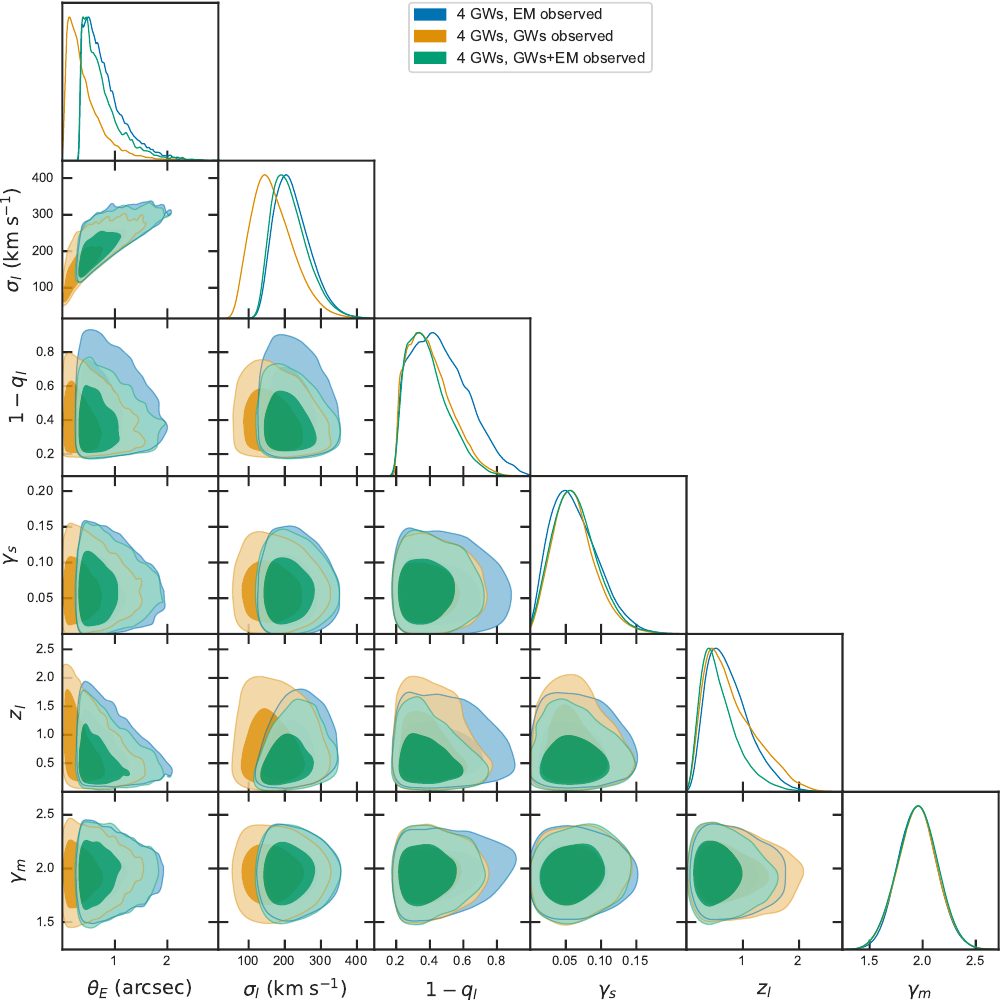}
    \caption{Lens mass model parameter distributions of the simulated sample of lens systems with quadruple images -- systems that could host a quadruply-lensed GW. Shown are the Einstein radius ($\theta_E$), lens galaxy velocity dispersion ($\sigma_l$), ellipticity ($1-q_l$), lens redshift ($z_l$), and the power-law index ($\gamma_m$) of the elliptical power-law model. The different colors refer to the different conditions on the lens systems: blue (\textit{EM deterved}) indicates a sample of lens systems as could be observed with \textit{Euclid}-like imaging, yellow (\textit{GWs observed}) indicates a sample of lens systems that are hosts of gravitational waves that have a high-enough SNR to be observed, and green (\textit{GWs+EM deterved}) indicates lens systems that are both observable with imaging and are host to a detectable gravitational wave, generated as described in \cref{sec:physrealisations}.}
    \label{fig:lenscorner}
\end{figure*}
\begin{table*}
    \centering
    \caption{All-sky lensing rates. Given are the number of lenses discoverable with \textit{Euclid} (assuming a \SI{15000}{\deg^2} survey size), and then for two scenarios the lensed gravitational wave rates, the percentage of the GW
lenses that correspond to an EM lens detectable with \textit{Euclid}, if it lies in its \SI{15000}{\deg^2} survey footprint, and the percentage of those EM lenses that could be identified correctly  -- defined as being the top hypothesis in our Bayesian evidence ranking (\cref{sec:identification}), and the percentage with an identification of over $2\sigma$ confidence over the next-most-likely hypothesis. We quote accuracies for the cases of an a-priori sky localisation of 1--5~\si{\deg^2} (e.g. 26.3--13.0 means an accuracy of 26.3\% in case of a \SI{1}{\deg^2} localisation, and 13.0\% for \SI{5}{\deg^2}.). The first scenario assumes that the galaxy's GW emission probability is $p(\text{GW emitted})\propto M_*$, and the second scenario assumes $p(\text{GW emitted})\propto\text{SFR}$. The three rows indicate the difference between requiring 2 or more, 3 or more, and 4 or more observable images/events. The effect of a different velocity dispersion function \citep{choiInternalCollectiveProperties2007} is shown in the lower three rows. We note that for this velocity dispersion function, we have not performed the end-to-end analysis to obtain identification accuracies, because it would not be a fair experiment: it only represents early-type galaxies, and given a gravitational wave, one cannot know a-priori that it came from an early-type galaxy.
    }
    \label{tab:rates}
\begin{tabular}{
  l
  r
  S[table-format=1.2e1]
  *{3}{%
   S[table-format=1.2]
   S[table-format=2.1]
   r
   r
  }
}
\toprule
{Vel. disp. function} & {Images} &  {\splitcell{EM lenses}} & \multicolumn{4}{c}{lensed GWs, $M_*$} & \multicolumn{4}{c}{lensed GWs, $\text{SFR}$}\\
& & & \si{yr^{-1}} & {\si{\percent} det.} & {\si{\percent} acc.} & {\si{\percent} $>2\sigma$} &\si{yr^{-1}} & {\si{\percent} det.} & {\si{\percent} acc.} & {\si{\percent} $>2\sigma$}\\
\midrule
\multirow{3}{*}{All galaxies} 
&\num{\ge2} & 3.25e5 & 0.47 & 24.5 & 26.3--13.0 & 11.0--5.3 & 5.05 &33.8 & 16.4--6.6 & 4.1--1.0 \\
&\num{\ge3} & 9.31e4 & 0.11 & 35.8 & 21.8--12.5 & 11.1--5.7 &1.15 & 43.8 & 29.8--14.9 & 11.5--3.6 \\
&\num{\ge4} & 8.29e4 &  0.04 & 39.4 & 52.4--39.2 & 36.8--23.0 &0.43 & 49.3 & 34.6--21.9 & 18.7--7.4 \\
\midrule
\multirow{3}{*}{Early-type galaxies}
&\num{\ge2} & 2.15e5 & 0.31 & 10.8 & & & 4.00 & 14.1 \\
&\num{\ge3} & 7.16e4 & 0.09 & 27.4 & & & 0.89 & 34.6 \\
&\num{\ge4} & 6.76e4 & 0.03 & 33.1 & & & 0.30 & 44.4 \\
\bottomrule
\end{tabular}
\end{table*}
With the lensed-GW population model as described in the previous section, we can generate a mock sample of lenses. This is split in two parts: in \cref{sec:physrealisations}, the physical parameters of the lenses are drawn, considering the lensing probabilities, as well as detectability criteria, and second, in \cref{sec:datagen}, a mock dataset is generated for each drawn physical realisation, with noise and uncertainties added.
\subsection{Physical realisation generation}
\label{sec:physrealisations}
The purpose of this paper is assessing our ability to link detected GWs with detected EM lenses. We therefore need 3 samples - one to analyse the properties of an EM-only lens sample, one to analyse the properties of a sample of detected lensed gravitational waves, and one to analyse a sample of systems where we can both detect the lens in the EM, as well as the lensed GW. We therefore run three different simulations, with different lensing probabilities:
\begin{enumerate}
    \item Simulation of EM lenses that \textit{Euclid} could detect in a random patch of sky
    \begin{equation}
    p_\text{lens}(\text{EM detected}) = p_\text{cross sec}P_\text{EMdet} \label{eq:plensemobs}
    \end{equation}
    \item Simulation of GW lenses that would emit a GW which will be detectable with LIGO-Virgo-Kagra 
    \begin{equation}
    p_\text{lens}(\text{GW observed}) = p_\text{cross sec}p_\text{emitted GW}P_\text{GWdet}\label{eq:plensgwobs}
    \end{equation}
    \item Simulation of EM+GW observed lenses that \textit{Euclid} would detect and that emit a GW which will be detectable with LIGO-Virgo-Kagra
    \begin{equation}
    p_\text{lens}(\text{EM + GW observed}) = p_\text{cross sec}p_\text{emitted GW}P_\text{GWdet}P_\text{EMdet}\label{eq:plensemgwobs}
    \end{equation}
\end{enumerate}

For each case, $\sim\!\!\!10^5$ physical realisations are sampled with the nested sampler \textit{Polychord}\footnote{Since Polychord and most other nested samplers do not deal well with plateaus in the likelihood \citep{fowlieNestedSamplingPlateaus2020}, in the code, strict plateaus are avoided.} \citep{handleyPolyChordNestedSampling2015,handleyPolyChordNextgenerationNested2015}. 
For the number of lenses and lensed GW rates, we have
\begin{align}
    N_\text{lenses} &= N(\text{lens galaxies}) \cdot N(\text{source galaxies}) \cdot \nonumber\\&\quad \expval{p_\text{lens}(\text{EM detected}|\vb*\theta)}_{\vb*\theta}\\
    &= \int\dd{z_l}\dd{\sigma_l}\pi_l(\sigma_l,z_l)\cdot N_\text{JAG}\frac{4\pi}{A_\text{JAG}}\cdot\nonumber\\&\quad\int\dd{\vb*\theta} p_\text{lens}(\text{EM detected}|\vb*\theta) \pi(\vb*\theta)\\
    N_\text{lensed GWs} &= \int\dd{z_l}\dd{\sigma_l}\pi_l(\sigma_l,z_l)\cdot N_\text{JAG}\frac{4\pi}{A_\text{JAG}}\cdot \nonumber\\&\quad\int\dd{\vb*\theta} p_\text{lens}(\text{GW detected}|\vb*\theta) \pi(\vb*\theta)
\end{align}
where $N(\text{lens/source galaxies})$ is the number of lens/source galaxies in the sky, $N_\text{JAG}$ is the number of galaxies in the Jaguar catalogue, $A_\text{JAG}$ is the sky area covered by the Jaguar catalogue, $\pi(\vb*\theta)$ is the normalised prior of the lens, source, and gravitational wave parameters, and $\expval{p_\text{lens}(\text{EM detected}|\vb*\theta)}_{\vb*\theta}$ is the probability for a random lens galaxy and a random source galaxy to be have parameters and positions such that it is a strong lens system, and that is also detectable in the EM. \textit{Polychord} is used to compute the integrals of the form $\int\dd{\vb*\theta} p_\text{lens}(\text{EM detected}|\vb*\theta) \pi(\vb*\theta)$, by making use of the fact that this integral is equivalent to the calculation of the evidence of a system with likelihood $p_\text{lens}(\text{EM detected}|\vb*\theta)$ and prior $\pi(\vb*\theta)$. \textit{Polychord} simultaneously also samples physical realisations of lens systems.
The distributions of the most important parameters of the lens mass model are shown in \cref{fig:lenscorner} and of the source galaxies are in \cref{fig:sourcecorner}. An overview of the predicted lensed GW rates and detected GW counts, as well as detectable fractions, is listed in \cref{tab:rates}.

We note that detectability is strongly correlated with large time delays, since large time delays are paired with larger and thus easier-to-observe lens systems. If a lensed gravitational wave is discovered, time delays are available without any modelling, and they provide a rough estimate of how probable they are to be observed in the EM, and how promising an electromagnetic follow-up could be. This is illustrated in \cref{fig:obs_time} for quads, where the fraction of EM-observable hosts rises with the time delay. Unfortunately, on the GW-side, the opposite is true --- larger time delays are harder to confidently pair to each other, because the false positive rate of pairing two unrelated GW events increases, so in reality there is a trade-off.

There is also a detection bias towards higher GW magnifications in SNR-limited samples. We discuss this shortly in \cref{app:magnification}.

\subsection{Mock data generation}
\label{sec:datagen}
\begin{figure*}
    \centering
    \includegraphics[width=\linewidth]{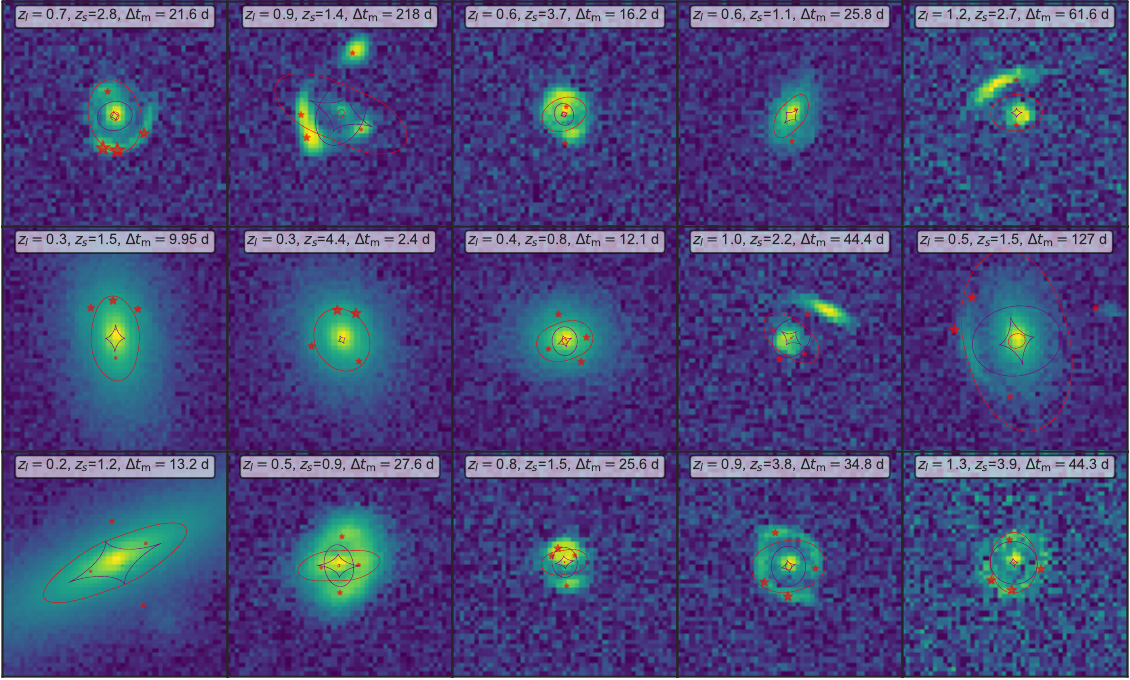}
    \caption{Examples of simulated lens systems. Systems in the top row were generated to have $\ge2$ images, systems in the middle row have $\ge4$ images, and systems in the bottom row are systems generated to be hosts of lensed GWs. A BBH source position is placed inside each source galaxy. Its source plane source position is indicated with the small green plus, its lensed source positions are indicated with the red stars, with the sizes representing the magnifications. The critical curves (red) and caustics (purple) are also shown. For each system, the lens redshift, source redshift and maximum time delay is listed.}
    \label{fig:panel}
\end{figure*}
For each of these physical realisations, a corresponding observed dataset is simulated. This includes
\begin{enumerate}
    \item Simulated \textit{Euclid}-like images were made with \texttt{lenstronomy}\footnote{\texttt{lenstronomy} has numerical issues with image calcuations for sources with very small S\'ersic radii, so in these cases, a point source approximation is made.} \citep{birrerGravitationalLensModeling2015,birrerLenstronomyIIGravitational2021,birrerLenstronomyMultipurposeGravitational2018}. \textit{Euclid} noise levels are described in \cref{sec:euclidsnr}. Some example images for the different physical realisations are shown in \cref{fig:panel}.
    \item A photometric redshift measurement is simulated, assuming a Gaussian error of $\Delta z/ z=0.05$  \cite[this approximately matches the expectations of][]{euclidcollaborationEuclidPreparationEuclid2020}.
\setcounter{nameOfYourChoice}{\value{enumi}}
\end{enumerate}
For the simulations of GW hosts (see \cref{sec:samplegen}), data derived from the GW signals is included:
\begin{enumerate}
\setcounter{enumi}{\value{nameOfYourChoice}}
    \item Uncertainties due to unmodelled substructure and model imperfections noises are added to the time delays and GW flux-ratios. Individual arrival times are given Gaussian noise of between 1.5-6\% of the largest time-delay. This conservative range is consistent with the expected time delay errors due to substructure \citep{liaoAnomaliesTimeDelays2018,gilmanTDCOSMOIIIDark2020}, and its value depends on the substructure mass spectrum and the line of sight environment. Individually measured magnifications are given log-normal uncertainties between 10-30\% (so relative magnifications have $\sqrt{2}$ times this error).\footnote{\citet{janquartFastPreciseMethodology2021}, for example, achieved a \textit{relative} magnification uncertainty of \SI{10}{\percent} for a high-signal event, and $\SI{\sim30}{\percent}$ for a subthreshold event.} Although exact numbers are unknown, for microlensing + substructure this is viewed as conservative when considering that GWs are not as susceptible to (stellar-mass) microlensing due to waveform suppression \citep{cheungStellarmassMicrolensingGravitational2021}. In reality, the scatter due to microlensing will be non-Gaussian, due to the non-linear nature of the substructure, and due to the magnification bias, but these effects are neglected in this work and considered secondary.
    \item The effective luminosity distance $d_L^{\rm eff}$ ($d_L/\mu^{1/2}$) is drawn from a log-normal distribution around the actual luminosity distance with $\sigma(\ln{d_L}) = 0.4$, which is approximately equal to the errors found on O1-O2 events \citep{abbottProspectsObservingLocalizing2020}. This is quite conservative, considering that when combining several GW events, the errors will reduce substantially.
    \item The sky localisation of the lensed GW event is drawn from a two-dimensional Gaussian, enclosing \SI{5}{deg^2} (\SI{90}{\percent}). In reality, the uncertainties depend on the orientation of detectors and the quality of the events themselves \citep{abbottProspectsObservingLocalizing2020}, so in \cref{sec:performance}, we vary this localisation precision over a wide range.
\end{enumerate}

\section{Gravitational Wave Event Localisation}
\label{sec:gwloc}
Using the simulations discussed in the previous section, we set up several mock identification runs. In \cref{sec:datalikelihood}, the likelihoods corresponding to the separate data products that are used in identification are explained. In \cref{sec:identification}, the identification process is described, and the expected performance in practice is assessed. Finally, in \cref{sec:subarcseclocalisation}, we investigate
the prospects for sub-arcsecond localisation of binary BHs via the
combination of all available data. 
\subsection{The likelihood}
\label{sec:datalikelihood}
For the identification and localisation, we use all of the mock data that was generated in \cref{sec:datagen}, which we split up in data from electromagnetic imaging, $\vb*y_\text{EM}$, consisting of the \textit{Euclid} imaging and (photometric) redshifts (part i-ii of \cref{sec:datagen}), and data from the gravitational waves, $\vb*y_\text{GW}$, consisting of flux-ratios, time-delays, an effective luminosity distance ($d_L/\mu^{1/2}$) measurement, and a sky localisation (part iii-vi of \cref{sec:datagen}). We similarly also have an associated data likelihood for EM part ($\mathcal{L}_\text{EM}=p(\vb*y_\text{EM}|\vb*\theta)$) and a GW part ($\mathcal{L}_\text{GW}=p(\vb*y_\text{GW}|\vb*\theta)$). All the likelihoods are constructed to match the generation of the data. The EM likelihood consists of
\begin{enumerate}
    \item The likelihood of the image, $\mathcal{L}_\text{img}(\vb*y_\text{EM} | \vb*\theta_i)$, is calculated by \texttt{lenstronomy}.
    \item The likelihood of the redshift is a Gaussian with uncertainty $\Delta z/z=0.05$. If $z_l>z_s$, then $\mathcal{L}_z=0$ (although it is unlikely, given the errors, one could measure a lens redshift higher than the source).
\setcounter{nameOfYourChoice}{\value{enumi}}
\end{enumerate}
The GW likelihood consists of the following:
\begin{enumerate}
\setcounter{enumi}{\value{nameOfYourChoice}}
    \item Although gravitational-wave detectors can resolve the arrival times to sub-millisecond accuracy, they can not be directly related to the lens model, because our lens model does not include substructure within the lens and along the line-of-sight, which can alter the arrival times. Therefore, we add to each arrival time a Gaussian error due to substructure with an uncertainty $\sigma_i$ of \SIlist{1.5;3;6}{\percent} of the largest time-delay. The uncertainty to account for modelling imprecisions due to substructure along the line of sight (the exceedingly small measurement errors are negligible). In reality, these errors between the images will be correlated and non-Gaussian, and could be absorbed in the lens model. However, uncorrelated Gaussian errors serve as a simple and conservative approximation. The distribution of the arrival times, constructed by the difference in the arrival times between the images, will have a non-diagonal covariance matrix. Using a time-delay likelihood with such a covariance matrix is equivalent to marginalizing over an unknown time offset with uncorrelated event time errors. 
    
    Explicitly, the $\log{\mathcal{L}}$ is:
    \begin{align}
        \log{\mathcal{L}}(\vb*{t}_d(\vb*{\theta})) = \mathcal{L}(\vb*{t}_d;\vb*{t}_{d,m},\vb*{\Sigma}_d),
    \end{align}
    where $\vb*{t}_d(\vb*{\theta})$ are the predicted time delays for model parameters $\vb*{\theta}$, $\vb*{t}_{d,m}$ are the measured time delays from the GW-side and $(\vb*{\Sigma}_d)_{ij} = \sigma_0^2\delta_{ij}+\sigma_0^2$ is the $3\times 3$ covariance matrix. This covariance matrix does have off-diagonal terms, and the derivation is given in \cref{app:margdelay}. The time delay uncertainty $\sigma_0$ is set to be \SIlist{1.5;3;6}{\percent} of the time between the first and last event.
    \item The flux ratio likelihood. The flux ratios are assumed to have log-normal errors of \SIlist{10;20;30}{\percent}. This error includes errors due microlensing and substructure along the line of sight as well as measurement errors from the GW-side. 
        The parities, or rather, the Morse phase indices of images \citep{daiWaveformsGravitationallyLensed2017,ezquiagaPhaseEffectsStrong2021,daiSearchLensedGravitational2020} are taken into account as well\footnote{We observe, however, only a marginal improvement in identification performance, \SI{<5}{\percent} better accuracies, when utilising this information. 
        }.
    The log-likelihood is set up analogously to the time-delay likelihood, with $\vb*{t}_d\to \ln(\frac{\vb*{F}[1...N]}{\vb*{F}[0]})$, and $\sigma_i\to \text{\SIlist{0.1;0.2;0.3}{}}$, when assuming \SIlist{10;20;30}{\percent} errors on the flux ratios. Note that the flux ratio and time-delay likelihood are linked: each image pairing is kept, and all permutations over images are taken as described in \cref{app:pairingimages}.
    \item The GW-inferred luminosity distance is modelled as a log-normal distribution, with $\sigma(\ln{d_L})=0.4$. In the future, this could be replaced by a likelihood function that more closely resembles the GW-inferred luminosity distance.
    \item The position (RA, DEC) information from the GW-side is assumed to be a two-dimensional Gaussian. The precise likelihood will depend on the sky localisation accuracy, which depends on orientation, the number of images of the event, and the GW SNRs. We assume that this likelihood is independent of any of the other parameters, and self-consistently assume a nominal localisation accuracy of \SI{5}{\deg^2} (90\%). 
\setcounter{nameOfYourChoice}{\value{enumi}}
\end{enumerate}

Lastly, because binary black holes formed from stellar remnants are expected to trace the star formation\footnote{In particular, the probability is proportional to its stellar mass or SFR, in line with the standard approach utilised in galaxy catalogue searches~\cite[e.g.][and references therein]{LIGOScientific:2019zcs}.}, 
we need to include this information in our model prior. Indeed, we have
\begin{enumerate}
\setcounter{enumi}{\value{nameOfYourChoice}}
    \item The astrophysical probability for a source galaxy to host a BBH (\cref{sec:lenssysprob}).
    \item The probability that (in the case of $n$ detected GWs) the BBH source position inside the source galaxy is within the $n$-image caustic, namely \cref{eq:pgwincaustic}.
\end{enumerate}
\subsection{Identification}
\label{sec:identification}
We now establish the formalism for the identification of the GWs. In \cref{sec:identificationsimple}, the identification is conceptually explained, in \cref{sec:idevidences}, the formal Bayesian method of identification is described, and in \cref{sec:performance}, the performance of the method on a realistic sample of lensed GWs is shown.
\subsubsection{Parameter overlaps}
\label{sec:identificationsimple}
\begin{figure}
    \centering
    \includegraphics[width=\linewidth]{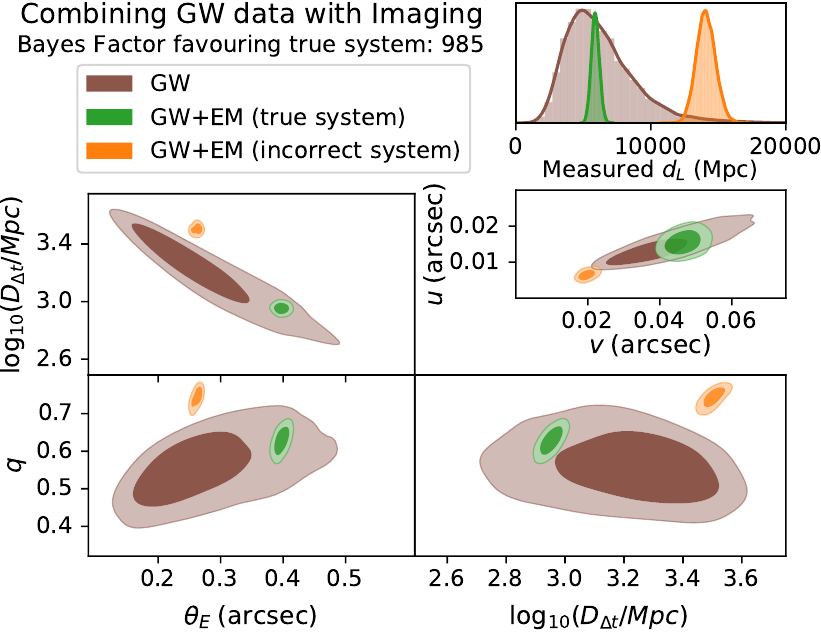}
    \caption{A simplified example of the identification process and joint modelling. A lensed GW event was generated according to the procedure outlined in the text, generating mock measured time delays, flux ratios, and a luminosity distance. This example assumes an isothermal density slope ($\gamma_l=2$), and no external shear. The posteriors of the inferred model parameters are shown in brown. In the top right panel, luminosity distance posteriors are shown. Additionally, a mock image and photometric redshift measurement from this galaxy was generated, and the posteriors of combining this EM data with the GW data are shown in green. For an unrelated lens, imaging and a photometric redshift measurement was also generated and combined with the measured GW, of which the posteriors are shown in orange. The top panel shows the luminosity distance measurement, and the bottom corner plot shows the posterior of the Einstein radius $\theta_E$, time-delay distance $D_{\Delta t}$ and axis ratio $q$. The small panel on the right shows the sub-arcsecond source BBH source position inference. Note that the true system overlaps better with the GW data than the incorrect system, which is also seen in the Bayes factor.}
    \label{fig:propfig}
\end{figure}

Conceptually, the identification of an EM lens system given the GW data amounts to determining which of the EM candidate lenses matches the GW data best in terms of model parameters. Given the GW data-set, a posterior of the lens and source model parameters is determined. This posterior is normalised and acts as prior on the model parameters that determine the evidence for the EM data for each lens system considered as potential lens. One way to intuitively visualise this is by doing the inference using the GW data on its own, and then comparing the resulting parameter posteriors to the posteriors of the candidate EM-lenses. For the GWs, there are the time delays, relative magnifications and a rough redshift estimate. The inference for a test system is shown in \cref{fig:propfig}. In this case, the inferred parameters overlap much better when using the EM data of the true host of the GW than an incorrect host, thus favouring the true system. Additionally, in the right panel it is shown that it is possible to identify the source position of the BBH to high precision in the panel in the middle right, another goal of the joint modelling. In the next section, a more formal evidence Bayesian ranking is introduced, which takes into account all priors and uses a more realistic model.

\subsubsection{Bayesian evidences}
\label{sec:idevidences}
Firstly, we assume that the detected dataset of lensed GW events, $\vb*y_\text{GW}$, which consists of multiple GW measurements that were identified as lensed signals of the same host. We assume that this identification was not a false positive. Folding in some false positive prior probability is straightforward, however, we cannot perform a population study on that in this work, because it is uncertain what the probability of falsely identifying a set of GWs as lensed from the GW-side. Furthermore, we assume that none of the \textit{Euclid}-identified lenses are false positives (generalisations are straightforward, but not expected to change the results).

From the GW-side, we have the data $\vb*y_\text{GW}$. From the EM-side, there is the data $\{\vb*y_{\text{EM},n}\}$, which consists of $N$ \textit{Euclid} images and (photometric) redshifts of lenses in a plausible area where the lensed GW event took place. The set of EM data of all observed EM-systems is denoted by $\{\vb*y_{\text{EM},n}\}$, where $\vb*y_{\text{EM},n}$ is the EM data of a single lens system $n$. Note: we assume that the plausible area on the sky is fully covered by \textit{Euclid} or a follow-up of \textit{Euclid}-like quality.

The Bayesian identification is a model comparison, where the different models are
\begin{enumerate}
    \item There is no electromagnetic imaging of the lens system that generated the lensed GW in our dataset $\{\vb*y_{\text{EM},n}\}$. This could be because the system is too faint or too small to be identified in imaging by \textit{Euclid} or with other data.
    \item There is electromagnetic imaging of the lens system that generated the lensed GW in our dataset $\{\vb*y_{\text{EM},n}\}$, and it came from the candidate lens system number~$i$.
\end{enumerate}

To each, there is a prior belief, denoted as $P(\text{not EM det})$ and $P(\text{EM det})$, which is the probability that any lensed GW has an detectable EM lens. These probabilities correspond to the detectable fractions as described in \cref{tab:rates}.

For model (i), the evidence is
\begin{align}
    \mathcal{Z}_\text{miss} &= P(\text{not EM det})\times p(\vb*y_\text{GW}|\text{not EM det})\times\prod_{j}p(\vb*y_\text{EM,j})\nonumber\\
    &=  P(\text{not EM det})\times\int\dd{\vb*\theta} p(\vb*y_\text{GW}|\vb*\theta)p(\vb*\theta|\text{not EM det})\nonumber\\&\quad\times\prod_{j}\int\dd{\vb*\theta_j}p(\vb*y_\text{EM,j}|\vb*\theta_j)p(\vb*\theta_j|\text{EM det})\nonumber\\
    &=  P(\text{not EM det})\times \int\dd{\vb*\theta} p(\vb*y_\text{GW}|\vb*\theta)p(\vb*\theta|\text{not EM det}) \times C\,,
\end{align}
that is, the model evidence encodes the probability that the system was not detected in the electromagnetic band
For model (ii), the evidence for image $i$ as host of the GW:
\begin{align}
    \mathcal{Z}_i &= \frac{P(\text{EM det})}{N_\text{lenses}}  \times p(\vb*y_\text{GW},\vb*y_\text{EM,i}|\text{EM det}) \times \prod_{j\neq i}p(\vb*y_\text{EM,j}|\text{EM det})\nonumber\\
    &=\frac{P(\text{EM det})}{N_\text{lenses}}  \times \int\dd{\vb*\theta_i}p(\vb*y_\text{GW},\vb*y_\text{EM,i}|\vb*\theta_i)p(\vb*\theta_i|\text{EM det}) \nonumber\\&\quad\times \prod_{j\neq i}\int\dd{\vb*\theta_j}p(\vb*y_\text{EM,j}|\vb*\theta_j)p(\vb*\theta_j|\text{EM det})\nonumber\\
    &=\frac{P(\text{EM det})}{N_\text{lenses}}  \times \frac{\int\dd{\vb*\theta_i}p(\vb*y_\text{GW},\vb*y_\text{EM,i}|\vb*\theta_i)p(\vb*\theta_i|\text{EM det})}{\int\dd{\vb*\theta_i}p(\vb*y_\text{EM,i}|\vb*\theta_i)p(\vb*\theta_i|\text{EM det})} \times C\nonumber\\
    &= \frac{P(\text{EM det})}{N_\text{lenses}} \times p(\vb*y_\text{GW}|\vb*y_\text{EM,i})\times C
\end{align}

The constant $C$ is not important as it is common to all of the evidences, and only the evidences $p(\vb*y_\text{GW}|\vb*y_\text{EM,i})$ are needed.
The factor $N_\text{lenses}$ is the number of detectable (i.e. findable in \textit{Euclid}-like images using modern lens-finding algorithms) lenses in the sky, and this factor is necessary to be able to compare the $\mathcal{Z}_i$'s with $\mathcal{Z}_\text{miss}$. 
One way to justify these factors is by considering the limit in which the likelihoods become non-informative ($\mathcal{L}_\text{GW}=\mathcal{L}_\text{EM}=1$). Then all integrals, which are just integrals over priors, go to $1$, and we are left with $\mathcal{Z}_\text{miss}=P(\text{not EM det})$ and $\mathcal{Z}_i=P(\text{EM det})/N_\text{lenses}$. 

To calculate the $\mathcal{Z}_i$'s, we rewrite the integrals to 
\begin{align}
p(\vb*y_\text{GW}|\vb*y_\text{EM,i}) &= \frac{\int\dd{\vb*\theta_i}\mathcal{L}_\text{GW} \mathcal{L}_\text{EM} p(\vb*\theta_i|\text{EM det})}{\int\dd{\vb*\theta_i}\mathcal{L}_\text{EM}p(\vb*\theta_i|\text{EM det})},
\end{align}
where the various $p(\vb*\theta_i|\ldots)$ serve as a prior on the lens system parameters. We self-consistently set these to be equal to the astrophysical priors, that also generated the simulated lens systems. This can be relaxed to less informative priors or be replaced with other astrophysically inferred priors. Although in the simulations, the lens galaxies were fixed at $n_l=4$, in the lens modelling, we do not use this, and instead model the lens S\'ersic index freely, with a wide uniform prior: $n_l\sim\mathcal{U}(0.5,7.5)$.
The evidence ratio is calculated by sampling the image-only information ($\mathcal{L}_\text{EM}\ p(\vb*\theta_i|\text{EM det})$), and then applying the importance sampling evidence estimator \citep{gewekeBayesianInferenceEconometric1989,nealAnnealedImportanceSampling2001} on the samples.

We chose not to include the detectability criteria (i.e. $p_\text{EM det}$ and $p_\text{GW det}$) in the evidence calculations, as they are not fundamental (we simply use hard cuts on the GW SNRs and detectability criteria), and it is more conservative to assess our achieved identification performances without them; this makes the prior more uninformative. Implicitly this removes the magnification bias, i.e. the bias stemming from the selection effect from detection thresholds, in the reconstruction. So it removes the restriction that the lens system and GWs must satisfy our detectibility requirements in our fitting. However, we note that for the magnification bias for EM galaxies (i.e. the omission of $p_\text{EM det}$), this is not a large issue, because any model that is compatible with the observed light distribution should anyway be classified as observable. The choice to ignore the GW magnification was largely made for computational reasons (which would not be an issue in reality), but the effect is not likely to be large because the luminosity distance likelihood already encompasses this information: the $p_\text{GW det}$ term would penalise models with GWs that are further away because they would be undetectable, but this information is largely already incorporated in the luminosity distance measurement.

\subsubsection{Performance}
\label{sec:performance}
\begin{figure*}
    \centering
    \includegraphics{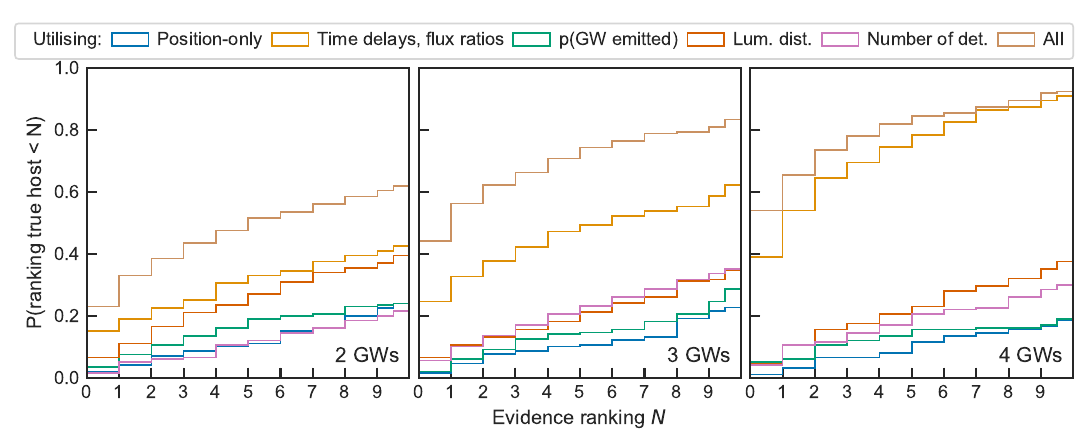}
    \caption[see figure]{A cumulative distribution function of the Bayesian rankings of the true system with respect to the most likely incorrect system. For example, for 2 GWs, the probability the true lens ranks first is \SI{23}{\percent}, and is in the top-5 candidates for \SI{48}{\percent} of the cases, when using combining all information. The three panels correspond to the case with 2, 3, and 4 detected GWs. The contribution of the different types of information used is split up into different colours, with in blue the position-only rankings. The other colors indicate the improved rankings upon considering also time delays and flux ratios (yellow), when considering the host GW emission probability (green), when considering the GW-inferred luminosity distance (orange-red), when considering the constraint that is obtained from the number of detected GWs (pink), and when combining all data (brown).
    We assume a $p(\text{GW emitted})\propto \text{SFR}$, a localisation area of \SI{5}{\deg^2}, and \SI{20}{\percent} and \SI{3}{\percent} errors for the flux ratios and time delays. 
   }
    \label{fig:rankings}
\end{figure*}
\begin{figure}
    \centering
    \includegraphics[width=\linewidth]{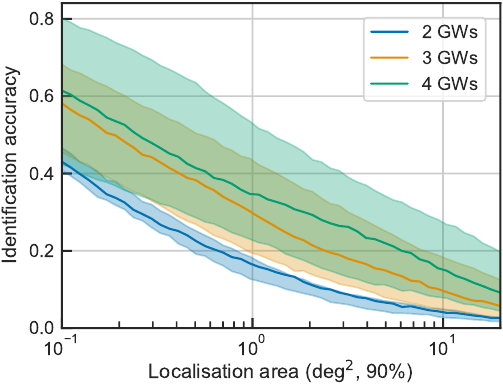}
    \caption{The identification accuracy of the sample with different localisation precisions. For the nominal scenario (the line), \SI{20}{\percent} and \SI{3}{\percent}  error on the fluxes and arrival times are assumed. The shaded region indicates the spread between the best case and the worst case scenario for flux ratio and time delay errors (\SI{10}{\percent} and \SI{1.5}{\percent} for the best case and \SI{30}{\percent} and \SI{6}{\percent} for the worst case).} 
    \label{fig:accuracylocalisation}
\end{figure}
To assess how reliable this Bayesian procedure works in identifying gravitational waves, what fraction we can identify accurately and to what confidences, a sample of 200 lensed GW events with their corresponding lens systems are (drawn from the distribution of \cref{eq:plensemgwobs}), and observed GW and EM data of them are generated. We focus on the model where the binary black hole population is linked to the star formation rate. For the BBH population linked to the old stellar population we performed the analysis as well, but results differ only slightly, with a slightly better identification performance for this model, as summarised in \cref{tab:rates}.
Additionally, around 400 normal (EM-only, so drawn from the distribution of \cref{eq:plensemobs}) lens systems\footnote{This 400 lenses means that given the \SI{20} lenses per \si{\deg^2} in Euclid we simulate, these lenses are enclosed in an area of about \SI{20}{\deg^2}. Including more lenses would not substantially change the results, because the probability of having the correct lens further than the 400th closest lens (thus being outside the \SI{\sim20}{\deg^2}) is very low (\num{<4e-4}). } were generated, with corresponding simulated EM data, to represent the closest systems that are not the true host. For each of the EM and the EM+GW systems, inference of the lens and the source model parameters on the EM data was done. Then, for each of the simulated GW datasets, and for each incorrect lens system (one of the 400 lens systems) as well as for the correct system that generated the GW, the evidences for generating the GW-dataset were calculated. For each GW dataset, this results in an evidence ranking, ordering the most likely EM lens highest, which ideally would be the true lens system. This way, a $200\times401$ grid is made, where for each of the GW datasets the evidences are calculated of the EM datasets. 
i

Because the sample of the 400 EM lens systems was selected based on position, the position likelihood is the most informative: it was already used to cut out the bulk of hundreds of thousands of lenses in the universe based on the \textit{Euclid} selection criterion. The performance is in fact most affected by the sky localisation from the GW-side. This is illustrated in \cref{fig:accuracylocalisation}, where we show the identification accuracy as a function of localisation precision. We define the identification accuracy as the fraction of gravitational waves that are identified as a top candidate, having the largest evidence of all candidates as well as having $\mathcal{Z}>\mathcal{Z}_\text{miss}$. To improve statistics, we oversample the positions, reporting the average accuracy over many realisations of the positions.

If the system is not ranked highest, it is still probably in the top candidates: the rank-order of the true system, with respect to the incorrect systems is shown in \cref{fig:rankings}. The different probabilities are split to assess the statistical power that each part of the data has. One can see that all components are informative, but high accuracies are only attained when combining all components: measured time delays and flux ratios, the astrophysical prior, and the luminosity distance likelihood are informative. This encourages such a joint analysis. 

\begin{figure*}
    \centering
    \includegraphics[width=\linewidth]{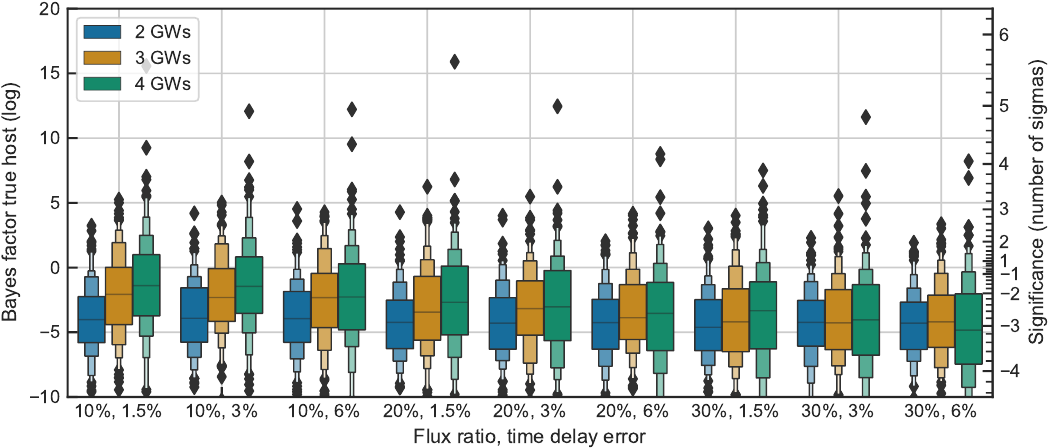}
    \caption{The identification Bayes factors for different numbers of GWs, for different uncertainties on GW fluxes and time delays. These are boxenplots: enhanced boxplots starting at the median, creating the central box between the first and third quartile, and from then each subsequent level contains half of the remaining data \citep{letter-value-plot}. The right axis indicates the sigma-equivalent to the Bayes factors on the left axis. We assume a sky localisation of \SI{5}{\deg^2}, and a GW emission probability $p(\text{GW emitted})\propto \text{SFR}$.}
    \label{fig:identificationpowerboxplots}
\end{figure*}

\begin{figure}
    \centering
    \includegraphics[width=\linewidth]{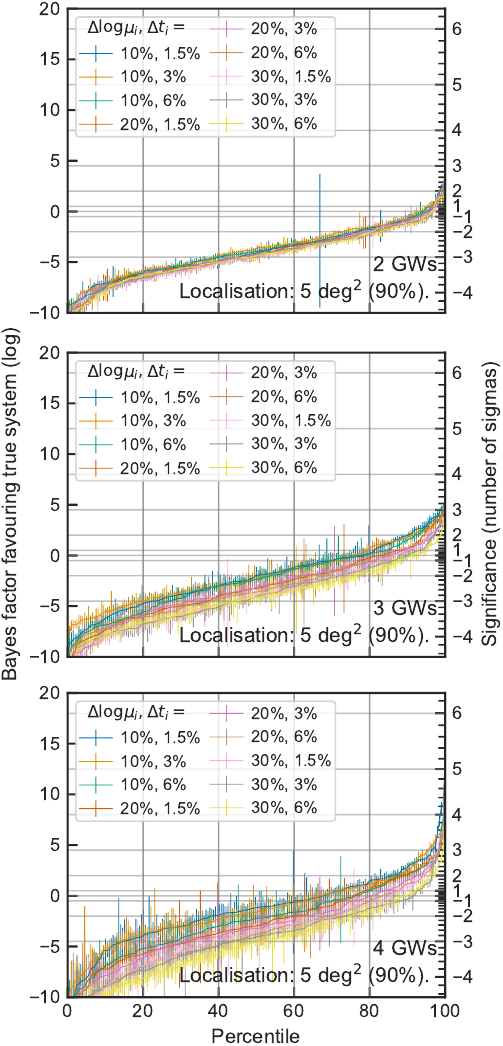}
    \caption{The identification confidences with estimated numerical uncertainties, given two, three and four detected GWs. Shown is the inverse CDF of the confidence level at which the GW is identified; the Bayes factor favouring it above the most likely injected lens, or if $\mathcal{Z}_\text{miss}$ is higher, the Bayes factor identifying it above $\mathcal{Z}_\text{miss}$. The legend indicates the uncertainty on the measured fluxes and on the measured times.}
    \label{fig:identificationpower}
\end{figure}

Whereas accuracies are simply derived from the rank-ordering of the lens systems, we also consider the identification Bayes factors with which the true lenses are identified over the most likely incorrect system. \cref{fig:identificationpowerboxplots} summarises these expected confidences of identification that we can expect, varying the number of detected GW events, and varying the uncertainties on the flux ratios and time delays. The full distributions, along with numerical uncertainties on the Bayes factors, are shown in \cref{fig:identificationpower}. We note that the accuracies are negatively impacted by numerical uncertainties. 

When doing such a probabilistic identification, there will inevitably be false positives. In the scenario for 4+ GWs, the true lens ranks top candidate in \SI{21.9}{\percent} of cases (see also \cref{tab:rates}), and a false lens ranks top in \SI{10}{\percent} of the cases. In the remaining \SI{68}{\percent} percent of cases, the hypothesis that the lens system was not observed was ranked with the highest evidence. The probability that the true lens ranks top with a $>\!2\sigma$ confidence is \SI{7.4}{\percent}, the probability that a false lens ranks top with $>\!2\sigma$ confidence is \SI{2.1}{\percent}.

\subsection{Sub-arcsecond localisation}
\label{sec:subarcseclocalisation}
\begin{figure*}
    \centering
    \includegraphics{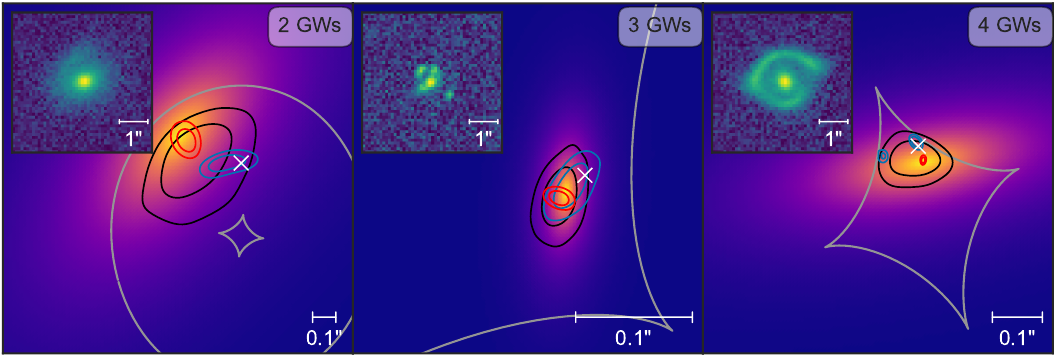}
    \caption{Sub-arcsecond localisations of GWs with lensing. Shown is the source plane light distribution, and overlayed the inference on the BBH position without any GW data (black) and the improvement when jointly modelling the EM+GW data (green-blue). The (red) contours for the galaxy centroid coordinates give an idea of the uncertainty in the EM modelling. The true BBH position (white x) is recovered. Additionally, the caustics are plotted (grey) and in the inset plots the observed lens images are displayed.}
    \label{fig:subarcseclocalisation}
\end{figure*}

In the case that a reliable host identification has been done, we can refine the source position of the BBH within the host galaxy by jointly modelling the EM+GW data. This is implicitly done in the identification framework (it is part of the evidence calculations), but we also separately investigate the localisation performance. We chose to look at the localisation for three typical systems, one system with two detected GWs, one with three, and one with four. The inferences of the BBH positions in the source plane, with respect to the host galaxy are shown in \cref{fig:subarcseclocalisation}. As expected, having more detected images will improve localisation. Although smaller source galaxies improve localisation in an absolute sense (due to the assumption that the BBH source positions follow the source light distribution), the downside is that the relative source position of the BBH within the galaxy is badly constrained, and moreover, smaller sources provide fewer constraints in the lens model and hence poorer identification of the EM system that hosts the GW.

In the right panel of \cref{fig:subarcseclocalisation}, the source position posterior is actually bimodal. This is caused by the symmetry of the lens model: for an elliptical power law lens, time delays, and magnifications remain constant upon reflection through the ellipses' major and minor axes. The source light distribution prior breaks this symmetry (we assume that the source position prior is proportional to the light intensity). External shear also breaks the symmetry (though not the 180-degree rotational symmetry), but generally the effect of an external shear is relatively small compared to uncertainties on magnifications and arrival times.

\section{Summary and discussion}
\label{sec:discussion}
We have assessed the possibility and developed a methodology for identifying the host galaxy of a lensed gravitational wave, and subsequently performing a sub-arcsecond localisation of a lensed binary black hole gravitational wave inside its host galaxy. Such identifications have the potential to answer important open questions in astrophysics, for example on the role and environment of binary black holes, on binary black hole physics, and time-delay cosmology. 

Considering a \textit{Euclid}-like instrument to identify lens systems and host galaxies, a GW event rate coupled to the young stellar population, four measured strongly-lensed gravitational waves, a combined sky localisation accuracy of \SI[parse-numbers=false]{\text{1-5}}{\deg^2}, nominal errors on the arrival times and magnifications (due to microlensing, substructure and measurement errors), we find that our evidence rankings identify the correct host as the top ranked candidate \SI[parse-numbers=false]{\text{34.6-21.9}}{\percent} of the cases, and for \SI[parse-numbers=false]{\text{18.7-7.3}}{\percent} of the cases, it is identified it with a Bayes factor $\ln(\mathcal{B})>2$ (this corresponds to a confidence of $\num{>2}\sigma$) above the next-highest ranked candidate. For only three (two) measured lensed GWs, while keeping the sky localisation precision constant, the accuracy drops to \SI[parse-numbers=false]{\text{11.5-3.6}}{\percent} (\SI[parse-numbers=false]{\text{4.1-1.0}}{\percent}). Subsequent localisation inside the host galaxy is also possible, and for well-measured systems very precise: this could be done to a precision of order \SI{10}{mas} for a bright and well-resolved system (\cref{fig:subarcseclocalisation}), but because of symmetries on the lens model and depending on the host galaxy's mass distribution, there may be multiple possible degenerate locations.

Additionally, we estimate the fraction of GW systems that is observable with \textit{Euclid}, and conservatively find that from \SIrange{20}{50}{\percent} of the lens systems generating detectable lensed GWs, depending on the assumptions on the velocity dispersion function and the number of detected GWs. The GW rate mainly depends on the assumed astrophysical GW environment, and because there is considerable uncertainty on the redshift evolution of the gravitational waves \citep{mukherjeeImpactAstrophysicalBinary2021}, we work out two scenarios, one scenario in which the GW population follows the old stellar population, and one scenario in which the GW population follows the star formation rate. The GW rate estimates based on a redshift evolution from stellar population models by \citet{wierdaDetectorHorizonForecasting2021} lie between this. The estimated number of lens systems observable with \textit{Euclid} is also very similar to the rate estimated by \citet{collettPopulationGalaxyGalaxyStrong2015}, who estimate \num{170000} detectable lenses in \textit{Euclid}'s footprint, versus our \num{215000} (that is, when assuming the same velocity dispersion function).

\subsection{Model limitations}
As with all simulations, simplifying approximations were made, some of which are conservative and some which are optimistic. 
For example, we treat all lenses as thin single-plane lenses, with a power-law and an external shear. 
We do not include an external convergence (a mass sheet) in the lens modelling. 
Although including this may make the lens modelling a bit harder, we note that similar degeneracies are already included in the form of an external shear and a free power-law index.  
Additionally, the assumption of normal errors on the arrival times and log-normal errors on the amplitudes is also an approximation. In reality, there will be correlations between the uncertainties, and the PDF of the amplitudes will be somewhat non-Gaussian: the magnifications and time-delays due to microlensing and line-of-sight substructures are non-Gaussian, and additionally, the magnification bias on the GW SNRs will not give rise to some non-Gaussian uncertainties (the SNRs determine the flux ratios). These are second-order effects however.
Furthermore, the lens and source light distribution was approximated as a S\'ersic ellipse. Although this approximation is acceptable for the lens galaxies, which are mostly massive ellipticals, especially for the larger source galaxies, this approximation is not that good. However, in the case of large source galaxies, the lens model is not a limiting factor to identification, and instead, we are limited by the data from the gravitational wave, while for small source galaxies, the PSF and pixel size of the imaging instrument will smoothen the source galaxies such that it is almost indistinguishable from a S\'ersic ellipse, so we do not think this will impact identification performance substantially.

In terms of estimating the rates, we have assumed a 100\% duty cycle for all detectors, but generally some time is spent 
either acquiring lock, unlocked and undergoing maintenance, unlocked due to unfavorable environmental conditions (earthquakes, wind, storms), or locked and in a state
of commissioning, where improvements are made to the
detectors. 
If a subset of the gravitational-wave detectors is online, the signal-to-noise ratio will be lower, which will degrade the rate of detectable lensed events. 
However, the duty cycle factor is being constantly improved (see e.g.,~\cite{aLIGO:2020wna} for improvements from 2015 to 2022), and we furthermore conservatively do not include the LIGO India detector, a fifth detector that is being built in India, which would improve the rate estimates. 
We have also assumed as a reference estimate a precise 5 deg$^2$ sky localisation, which is better than the current 2-detector and 3-detector sky localisation of typical O3 LIGO-Virgo-Kagra detections~\cite{LIGOScientific:2021djp}. 
A $\sim10$ deg$^2$ localisation is roughly indicative of an average sky localisation accuracy of a 4-detector or 5-detector network~\cite{Fairhurst:2017mvj}, worse than the $5$ deg$^2$ estimate. 
However, due to Earth's rotation, strong lensing would effectively increase the number of detectors~\cite[e.g.][]{Seto:2003iw,Goyal:2020bkm}; for example, in the case of a 4-image detection with 5 detectors online 100\% of the time, one would effectively have a network of close to 20 detectors, which would yield better sky localisation than a 5-detector network. 
Nevertheless, a systematic quantification with joint parameter estimation has not been done, and it requires folding in also duty cycles and low signal-to-noise ratio detections~\cite{Petrov:2021bqm}. 
Thus, a more detailed analysis regarding the duty cycles and sky localisations should be carried out in the future.

\subsection{Future prospects and follow-ups in the EM}
We also investigated the results of a ground-based follow-up with Subaru/HSC, as described in \cref{sec:groundbasedfollowup}, but found that $\SI{<1}{\percent}$ of lensed GWs could be identified correctly, suggesting that ground-based follow-up remains inefficient largely due to its limited spatial resolution. However, we only have done single-band modelling of lenses in this work, and the multiple colours that would be available from ground-based follow-ups would be helpful too, in both the lens detection and the lens modelling process. We leave this for future analyses. We also leave the analysis with follow-up observations with higher-resolution and more sensitive space-based instruments (e.g.\ Roman Space Telescope) for future analyses.

However, how likely is successful identification? 
This depends largely on the model assumptions. 
In the best-case scenario, where the astrophysical errors on time delays and flux ratios are minimal, and with a localisation of $\SI{1}{\deg^2}$, assuming a $\SI{40}{\percent}$ sky coverage by \textit{Euclid}, we can expect a $2\sigma$ identification with around $5$ quadruple image events, while in the nominal scenario for the errors, and \SI{5}{\deg^2} localisation, we would need around $13$ quadruple image events. 
Unfortunately, this seems to imply that the \textit{Euclid} survey will likely not be sufficient to make a successful identification.  

Instead, what would likely be required is a {\sl dedicated} follow-up from space of events outside the \textit{Euclid} footprint with similar or greater depth and resolution. 
A follow-up with similar quality to \textit{Euclid} could increase the success rate by at least a factor 2.5 (requiring a mere 2--5 quadruple image events) and possibly more with the upcoming Roman Space Telescope or even with the Hubble Space Telescope (a $\SI{1}{\deg^2}$ search area is smaller than the COSMOS field). 
Additionally, if the identification was inconclusive, follow-ups could still help. For instance, one could do a ground-based lens search that goes deeper than Euclid, which would pick up lenses that are too faint for \textit{Euclid}, and therefore lower increase our chances of finding a lens (and also decreasing the posterior probability in the hypothesis that the lens was not found, which helps with the confidence of identification).
However, we leave detailed estimates to future work. 
A lensed GW event would likely warrant such a dedicated follow-up program.

Note, also, that we have here assumed 2-sigma detection thresholds. 
If the threshold was set at 5-sigma, then the number of successful identifications would decrease significantly. 
However, if a 2-sigma or 3-sigma detection were made, it might warrant a more dedicated follow-up, which may allow us to discriminate the lens from others at higher accuracy. For instance, if a candidate has a badly constrained lens model, one could do a deeper or more high-resolution follow-up of this lens. The improvement in the precision of the model could then help us confirm or rule out the candidate with more confidence.
However, this, too, will require detailed estimates of the capabilities, which we leave to future work. 
We also note that even if the lens candidates are narrowed down to a few candidates, as opposed to one candidate, these systems can still contribute to statistical studies. 
For example, one might perform cosmography studies by marginalizing the Hubble constant measurement, weighted by the Bayes factor, for each candidate. 
By doing so, one might perform any follow-up science-case analysis in a statistical manner, similarly to the galaxy catalog methodologies developed for dark siren follow-up programs~\cite{LIGOScientific:2018gmd}.

\subsection{Improvements on lensed GW detections}
There are also other improvements that we may expect in the future. 
In particular, once built, a planned gravitational-wave detector, LIGO India, will increase the initial sky localisation accuracy from the gravitational wave. 
Furthermore, planned improvements to the lensing sub-threshold search pipelines~\cite{liFindingDiamondsRough2021,mcisaacSearchStronglyLensed2020} are expected to allow us to increase the number of quadruple image detections. 
Detection of more images, even if they have relatively low signal-to-noise ratios, would likely help with sky localisation by allowing for an effectively increased detector network in strong lensing analyses, contributing to narrowing down the a priori number of lensed candidates. 
A further improvement from the inclusion of sub-threshold events would be given by the time-delay (and to lesser degree magnification) information from the increased number of images; for example, four gravitational-wave images would pinpoint quite accurately the binary source position even without the magnification information, which would improve our ability to discriminate the correct lens candidate from the complete set of potential strong lens candidates.  
With these improvements, the number of events that are possible to localise in the coming years is expected to become more optimistic. 
Thus, the continued upgrades to the detector network and to the lensing pipelines will be crucial for successful lens identification. 

Finally, in the scenario that the current gravitational-wave detectors do not detect a sufficient number of lensed gravitational waves for the identification, the planned third-generation gravitational-wave detectors will. 
In particular, third-generation detectors such as the Einstein Telescope and Cosmic Explorer are expected to detect around $50$--$100$ lensed events per year~\citep{Biesiada:2014kwa,Li:2018prc,Xu:2021bfn}, some of which events are sure to be correctly localised to their host lens systems with a dedicated program.

A complementary avenue to localise not lensed binary black holes, but binary neutron stars, is the identification of counterparts to lensed binary neutron star events~\cite{Bianconi:2022etr, Smith:2022vbp}. 
In particular, if a strongly lensed binary neutron star is identified in low latency and its electromagnetic counterpart is located in a rapid follow-up search, this might likewise allow for identification of the host galaxy. 
Although here we target binary black holes, not neutron stars, it will still be interesting to consider this possibility, as many of the follow-up multimessenger applications, such as Hubble constant measurements, could also be performed with this type of search.

\subsection{Prospects for follow-up science}

If the lens system is uniquely identified, one gains access to the following information:
(i) Binary black hole and lens redshifts $(z_s, z_l)$ through spectroscopic or photometric measurements of the host and lens galaxies; 
(ii) Effective gravitational-wave luminosity distance $d_L^{\rm eff}$ (luminosity distance as measured directly from the gravitational wave, $d_L/\mu^{1/2}$)\footnote{Here and throughout we distinguish the luminosity distance as inferred from the gravitational waves without accounting for lensing ($d_L^{\rm eff}=d_L/\mu^{1/2}$) from the actual, intrinsic luminosity distance ($d_L$). }; (iii) Lens model reconstruction; (iv) Image properties (magnifications, time delays, Morse phases) as predicted by the lens model reconstruction and as directly inferred from the gravitational waves; (v) Precise gravitational-wave sky location; (vi) Host-galaxy properties (e.g.\ its stellar mass, age, metallicity), especially with additional follow-up observations;
(vii) Location of the binary black hole within the galaxy. Here we discuss potential applications enabled by this additional information.

Firstly, a measurement of (i) the source redshift and (ii) luminosity distance allows one to perform high-redshift standard-candle measurements~\citep{hannukselaLocalizingMergingBlack2020}. 
In particular, strongly lensed gravitational-wave events originate from redshifts $z_s \sim 1-4$, which are much above the typical binary neutron star redshifts $z_s\sim 0.01$~\citep{LIGOScientific:2017vwq}. 
This would allow us to probe cosmological expansion at a high redshift, a region currently beyond our reach with standard siren measurements. 
Moreover, as the gravitational wave at high redshift will have propagated an extensive distance, it could allow us to probe the gravitational-wave propagation in the context of alternative theories at much-improved accuracy, as shown by~\citet{finkeProbingModifiedGravitational2021}.

Secondly, access to the lens model reconstruction (iii) and the image properties from  (iv) could allow for improved lens modelling. 
A common hindrance to lens modelling is the mass-sheet degeneracy, which refers to the degeneracy between the power-law slope of the lens model and the lensing time delays. 
It has been shown, in the context of supernova lensing, that the degeneracy can be broken, once we obtain a standard-siren measurement of the system (iv), thus allowing for improved modelling of the lens itself~\citep{Oguri:2002ku}. 
In addition to being a standard siren, gravitational waves do not suffer from scattering, dispersion, or scintillation by the intergalactic and interstellar medium, and are less hampered by microlensing~\citep{cheungStellarmassMicrolensingGravitational2021}.

Thirdly, strongly lensed gravitational waves allow us to study the full polarization content of gravitational waves by increasing the effective number of detectors~\citep{Smith:2019polarisation,Goyal:2020bkm}. 
Once the sky location is known (v), one can further improve the polarization tests~\citep[e.g.][]{Pang:2020pfz}.

Fourthly, identifying the host galaxy (vi) and the merging black hole's location within it (vii) will allow us to study several interesting questions about the origin of merging black holes. 
For example, if they originate from galaxies, do they mostly belong to massive galaxies? Do they mainly come from old or young galaxies? Do they live in the bulges of galaxies, and do they follow the old stellar population, or in outskirts with more young galaxies?
Many different formation channels for binary black holes have been proposed \citep[and references therein]{santoliquidoCosmicMergerRate2021,mapelliBinaryBlackHole2020}, each having some dependence on the environment and redshift.
Locating the host galaxy of the binary black hole might allow us to investigate these questions in a way not possible with other methods. 
Moreover, precise, sub-arcsecond, GW localisation of these cosmological-distance events would allow us to study the problem under a microscope~\citep{hannukselaLocalizingMergingBlack2020}. 

Fifthly, owing to the localisation (vi), we could independently verify strong lensing detections~\citep{daiSearchLensedGravitational2020}. 
The searches for strong lensing rely on identifying two near-identical events. 
However, it is also plausible that two unrelated events appear identical within detector accuracy due to astrophysical coincidence~\citep{Caliskan:2022wbh}. 
An identification of the lens system in the electromagnetic band would allow us to independently confirm the detection.

\subsection{Future work}
In addition to what we have already incorporated in our identification methodology (\cref{sec:datalikelihood}), there is some additional information that can be used in practice that we have not included. For instance, we do not include the information given by the absence of detections in the full GW-databank: if some model parameters (or in extension, a candidate lens) predict an associated GW event with a high enough amplitude in a time window where no possible associated GW is seen (accounting for detector downtime and the response due to the antenna patterns of the detector), these parameters (lens) must be disfavoured. A similar line of thought was used by \citet{daiSearchLensedGravitational2020} to put an upper limit on the redshift of their candidate triplet of lensed GWs. 
Additionally, if we find that the gravitational wave is microlensed, this can give a wealth of extra information \citep{cheungStellarmassMicrolensingGravitational2021}.
Furthermore, a detailed analysis of the sky localisation capabilities for strongly lensed gravitational-waves with variable detector network is left for future work. 
This will be particularly interesting, as the strongly lensed gravitational-wave detections will allow for an effectively improved detector network, rendering potentially highly accurate sky localisations feasible \citep{Seto:2003iw}.

Future work could also generalise our results. For instance, in the lens modelling, the astrophysical priors (that were also used to generate the data) could be relaxed to more uninformative priors. This may make modelling the faint source galaxies a bit harder. One could also consider other instruments, for instance, the upcoming Chinese Space Station Telescope. The lensing simulations could also be made more realistic, for example by creating them by ray-tracing through cosmological simulations. Additionally, we have considered just the importance sampling evidence estimator. In some cases, the evidence uncertainty of this estimator is quite large. Although this could be solved by just sampling more points, more sophisticated evidence calculations, like annealed importance sampling \citep{nealAnnealedImportanceSampling2001} or nested sampling may be preferred. Of course, the disadvantage of such strategies is that the lens modelling cannot be done separately from the GW modelling: currently, the importance sampling estimator can simply be applied as a postprocessing step on the posterior samples of any third-party lens modelling software.

Lastly, although it was not originally intended for this purpose, the code can be easily used to calculate the astrophysical lensing probability for lenses just based on GW information too, because the lensing statistics prior is consistently folded in with the GW-data likelihood. For example, one could compute astrophysical contribution to evidence of the lensing hypothesis conditional on some set of GW events in our galaxy-galaxy lensing framework, even without any images of candidate hosts.

\section*{Acknowledgements}
We thank the referee for their careful reading and their suggestions that substantially improved the quality of this paper.
We would like to thank the Center for Information Technology of the University of Groningen for their support and for providing access to the Peregrine high performance computing cluster.
EW thanks Bharath Nagam, Bohdan Bidenko, and Thomas Collett for helpful discussions, and Simon Birrer for his support and for all his work on \texttt{lenstronomy}.
OAH thanks Thomas Collett, Tjonnie Li, Anupreeta More, Narola Harsh, Haris K, and the LVC lensing group for useful discussions. O.A.H. acknowledge support by grants from the Research Grants Council of Hong Kong (Project No. CUHK 14304622 and 14307923), the start-up grant from the Chinese University of Hong Kong, and the Direct Grant for Research from the Research Committee of The Chinese University of Hong Kong. 
We would also like to thank Imre Bartos for useful discussion on terminology. 

\section*{Data Availability}

The code to run all simulations, to generate the tables and figures, is made available on reasonable request from EW.

\bibliographystyle{mnras}
\bibliography{ref,references} %

\begin{thebibliography}{}
\makeatletter
\relax
\def\mn@urlcharsother{\let\do\@makeother \do\$\do\&\do\#\do\^\do\_\do\%\do\~}
\def\mn@doi{\begingroup\mn@urlcharsother \@ifnextchar [ {\mn@doi@}
  {\mn@doi@[]}}
\def\mn@doi@[#1]#2{\def\@tempa{#1}\ifx\@tempa\@empty \href
  {http://dx.doi.org/#2} {doi:#2}\else \href {http://dx.doi.org/#2} {#1}\fi
  \endgroup}
\def\mn@eprint#1#2{\mn@eprint@#1:#2::\@nil}
\def\mn@eprint@arXiv#1{\href {http://arxiv.org/abs/#1} {{\tt arXiv:#1}}}
\def\mn@eprint@dblp#1{\href {http://dblp.uni-trier.de/rec/bibtex/#1.xml}
  {dblp:#1}}
\def\mn@eprint@#1:#2:#3:#4\@nil{\def\@tempa {#1}\def\@tempb {#2}\def\@tempc
  {#3}\ifx \@tempc \@empty \let \@tempc \@tempb \let \@tempb \@tempa \fi \ifx
  \@tempb \@empty \def\@tempb {arXiv}\fi \@ifundefined
  {mn@eprint@\@tempb}{\@tempb:\@tempc}{\expandafter \expandafter \csname
  mn@eprint@\@tempb\endcsname \expandafter{\@tempc}}}

\bibitem[\protect\citeauthoryear{Aasi et~al.}{Aasi
  et~al.}{2015}]{LIGOScientific:2014pky}
Aasi J.,  et~al., 2015, \mn@doi [Class. Quant. Grav.]
  {10.1088/0264-9381/32/7/074001}, 32, 074001

\bibitem[\protect\citeauthoryear{Abbott et~al.}{Abbott
  et~al.}{2017}]{LIGOScientific:2017vwq}
Abbott B.~P.,  et~al., 2017, \mn@doi [Phys. Rev. Lett.]
  {10.1103/PhysRevLett.119.161101}, 119, 161101

\bibitem[\protect\citeauthoryear{Abbott et~al.}{Abbott
  et~al.}{2019}]{LIGOScientific:2018mvr}
Abbott B.~P.,  et~al., 2019, \mn@doi [Phys. Rev. X]
  {10.1103/PhysRevX.9.031040}, 9, 031040

\bibitem[\protect\citeauthoryear{Abbott et~al.,}{Abbott
  et~al.}{2020a}]{abbottProspectsObservingLocalizing2020}
Abbott B.~P.,  et~al., 2020a, \mn@doi [Living Rev. Relativ.]
  {10.1007/s41114-020-00026-9}, 23, 3

\bibitem[\protect\citeauthoryear{Abbott et~al.}{Abbott
  et~al.}{2020b}]{Abbott:2020qfu}
Abbott B.~P.,  et~al., 2020b, \mn@doi [Living Rev. Relativ.]
  {10.1007/s41114-020-00026-9}, 23, 3

\bibitem[\protect\citeauthoryear{Abbott et~al.}{Abbott
  et~al.}{2021a}]{LIGOScientific:2020ibl}
Abbott R.,  et~al., 2021a, \mn@doi [Phys. Rev. X] {10.1103/PhysRevX.11.021053},
  11, 021053

\bibitem[\protect\citeauthoryear{Abbott et~al.}{Abbott
  et~al.}{2021b}]{LIGOScientific:2019zcs}
Abbott B.~P.,  et~al., 2021b, \mn@doi [Astrophys. J.]
  {10.3847/1538-4357/abdcb7}, 909, 218

\bibitem[\protect\citeauthoryear{Abbott et~al.,}{Abbott
  et~al.}{2021c}]{abbottPopulationPropertiesCompact2021}
Abbott R.,  et~al., 2021c, \mn@doi [ApJ] {10.3847/2041-8213/abe949}, 913, L7

\bibitem[\protect\citeauthoryear{Abbott et~al.,}{Abbott
  et~al.}{2021d}]{theligoscientificcollaborationSearchLensingSignatures2021}
Abbott R.,  et~al., 2021d, arXiv e-prints, 2105, arXiv:2105.06384

\bibitem[\protect\citeauthoryear{Acernese et~al.}{Acernese
  et~al.}{2015}]{VIRGO:2014yos}
Acernese F.,  et~al., 2015, \mn@doi [Class. Quant. Grav.]
  {10.1088/0264-9381/32/2/024001}, 32, 024001

\bibitem[\protect\citeauthoryear{Aihara et~al.,}{Aihara
  et~al.}{2018}]{aiharaFirstDataRelease2018}
Aihara H.,  et~al., 2018, \mn@doi [PASJ] {10.1093/pasj/psx081}, 70, S8

\bibitem[\protect\citeauthoryear{Akutsu et~al.,}{Akutsu
  et~al.}{2020}]{Akutsu:2020his}
Akutsu T.,  et~al., 2020, ArXiv200505574 Astro-Ph Physicsgr-Qc Physicsphysics

\bibitem[\protect\citeauthoryear{Artale, Mapelli, Giacobbo, Sabha, Spera,
  Santoliquido  \& Bressan}{Artale
  et~al.}{2019}]{artaleHostGalaxiesMerging2019}
Artale M.~C.,  Mapelli M.,  Giacobbo N.,  Sabha N.~B.,  Spera M.,  Santoliquido
  F.,   Bressan A.,  2019, \mn@doi [MNRAS] {10.1093/mnras/stz1382}, 487, 1675

\bibitem[\protect\citeauthoryear{Aso, Michimura, Somiya, Ando, Miyakawa,
  Sekiguchi, Tatsumi  \& Yamamoto}{Aso et~al.}{2013}]{Aso:2013eba}
Aso Y.,  Michimura Y.,  Somiya K.,  Ando M.,  Miyakawa O.,  Sekiguchi T.,
  Tatsumi D.,   Yamamoto H.,  2013, \mn@doi [Phys. Rev. D]
  {10.1103/PhysRevD.88.043007}, 88, 043007

\bibitem[\protect\citeauthoryear{Auger, Treu, Bolton, Gavazzi, Koopmans,
  Marshall, Moustakas  \& Burles}{Auger et~al.}{2010}]{augerSloanLensACS2010}
Auger M.~W.,  Treu T.,  Bolton A.~S.,  Gavazzi R.,  Koopmans L. V.~E.,
  Marshall P.~J.,  Moustakas L.~A.,   Burles S.,  2010, \mn@doi [ApJ]
  {10.1088/0004-637X/724/1/511}, 724, 511

\bibitem[\protect\citeauthoryear{Barkana}{Barkana}{1998}]{barkanaFastCalculationFamily1998}
Barkana R.,  1998, \mn@doi [ApJ] {10.1086/305950}, 502, 531

\bibitem[\protect\citeauthoryear{Bernardi et~al.,}{Bernardi
  et~al.}{2003}]{bernardiEarlyTypeGalaxiesSloan2003}
Bernardi M.,  et~al., 2003, \mn@doi [AJ] {10.1086/367794}, 125, 1866

\bibitem[\protect\citeauthoryear{Bernardi, Shankar, Hyde, Mei, Marulli  \&
  Sheth}{Bernardi et~al.}{2010}]{bernardiGalaxyLuminositiesStellar2010}
Bernardi M.,  Shankar F.,  Hyde J.~B.,  Mei S.,  Marulli F.,   Sheth R.~K.,
  2010, \mn@doi [MNRAS] {10.1111/j.1365-2966.2010.16425.x}, 404, 2087

\bibitem[\protect\citeauthoryear{Bianconi et~al.,}{Bianconi
  et~al.}{2022}]{Bianconi:2022etr}
Bianconi M.,  et~al., 2022, arXiv e-prints

\bibitem[\protect\citeauthoryear{Biesiada, Ding, Piorkowska  \& Zhu}{Biesiada
  et~al.}{2014}]{Biesiada:2014kwa}
Biesiada M.,  Ding X.,  Piorkowska A.,   Zhu Z.-H.,  2014, \mn@doi [JCAP]
  {10.1088/1475-7516/2014/10/080}, 10, 080

\bibitem[\protect\citeauthoryear{Bigdeli, Lin, Portenier, Dunbar  \&
  Zwicker}{Bigdeli et~al.}{2020}]{bigdeliLearningGenerativeModels2020}
Bigdeli S.~A.,  Lin G.,  Portenier T.,  Dunbar L.~A.,   Zwicker M.,  2020,
  arXiv e-prints, p. arXiv:2001.02728

\bibitem[\protect\citeauthoryear{Birrer \& Amara}{Birrer \&
  Amara}{2018}]{birrerLenstronomyMultipurposeGravitational2018}
Birrer S.,  Amara A.,  2018, \mn@doi [Physics of the Dark Universe]
  {10.1016/j.dark.2018.11.002}, 22, 189

\bibitem[\protect\citeauthoryear{Birrer, Amara  \& Refregier}{Birrer
  et~al.}{2015}]{birrerGravitationalLensModeling2015}
Birrer S.,  Amara A.,   Refregier A.,  2015, \mn@doi [ApJ]
  {10.1088/0004-637X/813/2/102}, 813, 102

\bibitem[\protect\citeauthoryear{Birrer et~al.,}{Birrer
  et~al.}{2021}]{birrerLenstronomyIIGravitational2021}
Birrer S.,  et~al., 2021, \mn@doi [The Journal of Open Source Software]
  {10.21105/joss.03283}, 6, 3283

\bibitem[\protect\citeauthoryear{Bolton, Treu, Koopmans, Gavazzi, Moustakas,
  Burles, Schlegel  \& Wayth}{Bolton et~al.}{2008}]{boltonSloanLensACS2008}
Bolton A.~S.,  Treu T.,  Koopmans L. V.~E.,  Gavazzi R.,  Moustakas L.~A.,
  Burles S.,  Schlegel D.~J.,   Wayth R.,  2008, \mn@doi [ApJ]
  {10.1086/589989}, 684, 248

\bibitem[\protect\citeauthoryear{Buikema et~al.}{Buikema
  et~al.}{2020}]{aLIGO:2020wna}
Buikema A.,  et~al., 2020, \mn@doi [Phys. Rev. D]
  {10.1103/PhysRevD.102.062003}, 102, 062003

\bibitem[\protect\citeauthoryear{{\c C}al{\i}{\c s}kan, Ezquiaga, Hannuksela
  \& Holz}{{\c C}al{\i}{\c s}kan et~al.}{2022}]{Caliskan:2022wbh}
{\c C}al{\i}{\c s}kan M.,  Ezquiaga J.~M.,  Hannuksela O.~A.,   Holz D.~E.,
  2022, ArXiv220104619 Astro-Ph Physicsgr-Qc

\bibitem[\protect\citeauthoryear{Cao, Li  \& Wang}{Cao
  et~al.}{2014}]{Cao:2014oaa}
Cao Z.,  Li L.-F.,   Wang Y.,  2014, \mn@doi [Phys. Rev. D]
  {10.1103/PhysRevD.90.062003}, 90, 062003

\bibitem[\protect\citeauthoryear{Cheung, Gais, Hannuksela  \& Li}{Cheung
  et~al.}{2021}]{cheungStellarmassMicrolensingGravitational2021}
Cheung M. H.~Y.,  Gais J.,  Hannuksela O.~A.,   Li T. G.~F.,  2021, \mn@doi
  [MNRAS] {10.1093/mnras/stab579}, 503, 3326

\bibitem[\protect\citeauthoryear{Choi, Park  \& Vogeley}{Choi
  et~al.}{2007}]{choiInternalCollectiveProperties2007}
Choi Y.-Y.,  Park C.,   Vogeley M.~S.,  2007, \mn@doi [ApJ] {10.1086/511060},
  658, 884

\bibitem[\protect\citeauthoryear{Christian, Vitale  \& Loeb}{Christian
  et~al.}{2018}]{Christian:2018vsi}
Christian P.,  Vitale S.,   Loeb A.,  2018, \mn@doi [Phys. Rev. D]
  {10.1103/PhysRevD.98.103022}, 98, 103022

\bibitem[\protect\citeauthoryear{Collett}{Collett}{2015}]{collettPopulationGalaxyGalaxyStrong2015}
Collett T.~E.,  2015, \mn@doi [ApJ] {10.1088/0004-637X/811/1/20}, 811, 20

\bibitem[\protect\citeauthoryear{Cropper et~al.,}{Cropper
  et~al.}{2016}]{cropperVISVisibleImager2016}
Cropper M.,  et~al., 2016, in MacEwen H.~A.,  Fazio G.~G.,  Lystrup M.,
  Batalha N.,  Siegler N.,   Tong E.~C.,  eds,  Society of {{Photo-Optical
  Instrumentation Engineers}} ({{SPIE}}) {{Conference Series}} Vol. 9904, Space
  {{Telescopes}} and {{Instrumentation}} 2016: {{Optical}}, {{Infrared}}, and
  {{Millimeter Wave}}. p. 99040Q, \mn@doi{10.1117/12.2234739}

\bibitem[\protect\citeauthoryear{Dai \& Venumadhav}{Dai \&
  Venumadhav}{2017}]{daiWaveformsGravitationallyLensed2017}
Dai L.,  Venumadhav T.,  2017, arXiv e-prints, p. arXiv:1702.04724

\bibitem[\protect\citeauthoryear{Dai, Zackay, Venumadhav, Roulet  \&
  Zaldarriaga}{Dai et~al.}{2020}]{daiSearchLensedGravitational2020}
Dai L.,  Zackay B.,  Venumadhav T.,  Roulet J.,   Zaldarriaga M.,  2020, arXiv
  e-prints, \href {https://ui.adsabs.harvard.edu/abs/2020arXiv200712709D} {p.
  arXiv:2007.12709}

\bibitem[\protect\citeauthoryear{{Etherington} et~al.,}{{Etherington}
  et~al.}{2023}]{etheringtonExternalShearIsNotShear}
{Etherington} A.,  et~al., 2023, \mn@doi [arXiv e-prints]
  {10.48550/arXiv.2301.05244}, \href
  {https://ui.adsabs.harvard.edu/abs/2023arXiv230105244E} {p. arXiv:2301.05244}

\bibitem[\protect\citeauthoryear{{Euclid Collaboration} et~al.,}{{Euclid
  Collaboration} et~al.}{2019}]{euclidcollaborationEuclidPreparationIV2019}
{Euclid Collaboration} et~al., 2019, \mn@doi [A\&A]
  {10.1051/0004-6361/201935187}, 627, A59

\bibitem[\protect\citeauthoryear{{Euclid Collaboration} et~al.,}{{Euclid
  Collaboration} et~al.}{2020}]{euclidcollaborationEuclidPreparationEuclid2020}
{Euclid Collaboration} et~al., 2020, \mn@doi [A\&A]
  {10.1051/0004-6361/202039403}, 644, A31

\bibitem[\protect\citeauthoryear{Ezquiaga, Holz, Hu, Lagos  \& Wald}{Ezquiaga
  et~al.}{2021}]{ezquiagaPhaseEffectsStrong2021}
Ezquiaga J.~M.,  Holz D.~E.,  Hu W.,  Lagos M.,   Wald R.~M.,  2021, \mn@doi
  [Phys. Rev. D] {10.1103/PhysRevD.103.064047}, 103, 064047

\bibitem[\protect\citeauthoryear{Fairhurst}{Fairhurst}{2018}]{Fairhurst:2017mvj}
Fairhurst S.,  2018, \mn@doi [Class. Quant. Grav.] {10.1088/1361-6382/aab675},
  35, 105002

\bibitem[\protect\citeauthoryear{F{\'e}ron, Hjorth, McKean  \&
  Samsing}{F{\'e}ron et~al.}{2009}]{feronSearchDiskGalaxyLenses2009}
F{\'e}ron C.,  Hjorth J.,  McKean J.~P.,   Samsing J.,  2009, \mn@doi [ApJ]
  {10.1088/0004-637X/696/2/1319}, 696, 1319

\bibitem[\protect\citeauthoryear{Finke, Foffa, Iacovelli, Maggiore  \&
  Mancarella}{Finke et~al.}{2021}]{finkeProbingModifiedGravitational2021}
Finke A.,  Foffa S.,  Iacovelli F.,  Maggiore M.,   Mancarella M.,  2021, arXiv
  e-prints, p. arXiv:2107.05046

\bibitem[\protect\citeauthoryear{Fishbach et~al.}{Fishbach
  et~al.}{2019}]{LIGOScientific:2018gmd}
Fishbach M.,  et~al., 2019, \mn@doi [Astrophys. J. Lett.]
  {10.3847/2041-8213/aaf96e}, 871, L13

\bibitem[\protect\citeauthoryear{Fowlie, Handley  \& Su}{Fowlie
  et~al.}{2020}]{fowlieNestedSamplingPlateaus2020}
Fowlie A.,  Handley W.,   Su L.,  2020, arXiv e-prints, 2010, arXiv:2010.13884

\bibitem[\protect\citeauthoryear{Genel et~al.,}{Genel
  et~al.}{2014}]{genelIntroducingIllustrisProject2014}
Genel S.,  et~al., 2014, \mn@doi [MNRAS] {10.1093/mnras/stu1654}, 445, 175

\bibitem[\protect\citeauthoryear{Geweke}{Geweke}{1989}]{gewekeBayesianInferenceEconometric1989}
Geweke J.,  1989, \mn@doi [Econometrica] {10.2307/1913710}, 57, 1317

\bibitem[\protect\citeauthoryear{Gilman, Birrer  \& Treu}{Gilman
  et~al.}{2020}]{gilmanTDCOSMOIIIDark2020}
Gilman D.,  Birrer S.,   Treu T.,  2020, \mn@doi [A\&A]
  {10.1051/0004-6361/202038829}, 642, A194

\bibitem[\protect\citeauthoryear{Goldstein, Nugent  \& Goobar}{Goldstein
  et~al.}{2019}]{goldsteinRatesPropertiesSupernovae2019}
Goldstein D.~A.,  Nugent P.~E.,   Goobar A.,  2019, \mn@doi [ApJS]
  {10.3847/1538-4365/ab1fe0}, 243, 6

\bibitem[\protect\citeauthoryear{Gong et~al.,}{Gong
  et~al.}{2019}]{gongCosmologyChineseSpace2019}
Gong Y.,  et~al., 2019, \mn@doi [ApJ] {10.3847/1538-4357/ab391e}, 883, 203

\bibitem[\protect\citeauthoryear{Goyal, Haris, Mehta  \& Ajith}{Goyal
  et~al.}{2021a}]{Goyal:2020bkm}
Goyal S.,  Haris K.,  Mehta A.~K.,   Ajith P.,  2021a, \mn@doi [Phys. Rev. D]
  {10.1103/PhysRevD.103.024038}, 103, 024038

\bibitem[\protect\citeauthoryear{Goyal, D., Kapadia  \& Ajith}{Goyal
  et~al.}{2021b}]{Goyal:2021hxv}
Goyal S.,  D. H.,  Kapadia S.~J.,   Ajith P.,  2021b, \mn@doi [Phys. Rev. D]
  {10.1103/PhysRevD.104.124057}, 104, 124057

\bibitem[\protect\citeauthoryear{Handley, Hobson  \& Lasenby}{Handley
  et~al.}{2015a}]{handleyPolyChordNestedSampling2015}
Handley W.~J.,  Hobson M.~P.,   Lasenby A.~N.,  2015a, \mn@doi [MNRAS Letters]
  {10.1093/mnrasl/slv047}, 450, L61

\bibitem[\protect\citeauthoryear{Handley, Hobson  \& Lasenby}{Handley
  et~al.}{2015b}]{handleyPolyChordNextgenerationNested2015}
Handley W.~J.,  Hobson M.~P.,   Lasenby A.~N.,  2015b, \mn@doi [MNRAS]
  {10.1093/mnras/stv1911}, 453, 4385

\bibitem[\protect\citeauthoryear{Hannuksela, Haris, Ng, Kumar, Mehta, Keitel,
  Li  \& Ajith}{Hannuksela
  et~al.}{2019}]{hannukselaSearchGravitationalLensing2019}
Hannuksela O.~A.,  Haris K.,  Ng K. K.~Y.,  Kumar S.,  Mehta A.~K.,  Keitel D.,
   Li T. G.~F.,   Ajith P.,  2019, \mn@doi [ApJ] {10.3847/2041-8213/ab0c0f},
  874, L2

\bibitem[\protect\citeauthoryear{Hannuksela, Collett, {\c C}al{\i}{\c s}kan  \&
  Li}{Hannuksela et~al.}{2020}]{hannukselaLocalizingMergingBlack2020}
Hannuksela O.~A.,  Collett T.~E.,  {\c C}al{\i}{\c s}kan M.,   Li T. G.~F.,
  2020, \mn@doi [MNRAS] {10.1093/mnras/staa2577}, 498, 3395

\bibitem[\protect\citeauthoryear{Haris, Mehta, Kumar, Venumadhav  \&
  Ajith}{Haris et~al.}{2018}]{harisIdentifyingStronglyLensed2018}
Haris K.,  Mehta A.~K.,  Kumar S.,  Venumadhav T.,   Ajith P.,  2018, arXiv
  e-prints, 1807, arXiv:1807.07062

\bibitem[\protect\citeauthoryear{Harry}{Harry}{2010}]{Harry:2010zz}
Harry G.~M.,  2010, \mn@doi [Class. Quant. Grav.]
  {10.1088/0264-9381/27/8/084006}, 27, 084006

\bibitem[\protect\citeauthoryear{Hartley, Flamary, Jackson, Tagore  \&
  Metcalf}{Hartley et~al.}{2017}]{hartleySupportVectorMachine2017}
Hartley P.,  Flamary R.,  Jackson N.,  Tagore A.~S.,   Metcalf R.~B.,  2017,
  \mn@doi [MNRAS] {10.1093/mnras/stx1733}, 471, 3378

\bibitem[\protect\citeauthoryear{Henriques, White, Thomas, Angulo, Guo, Lemson,
  Springel  \& Overzier}{Henriques
  et~al.}{2015}]{henriquesGalaxyFormationPlanck2015}
Henriques B. M.~B.,  White S. D.~M.,  Thomas P.~A.,  Angulo R.,  Guo Q.,
  Lemson G.,  Springel V.,   Overzier R.,  2015, \mn@doi [MNRAS]
  {10.1093/mnras/stv705}, 451, 2663

\bibitem[\protect\citeauthoryear{Hofmann, Kafadar  \& Wickham}{Hofmann
  et~al.}{2011}]{letter-value-plot}
Hofmann H.,  Kafadar K.,   Wickham H.,  2011, Technical report, Letter-Value
  Plots: {{Boxplots}} for Large Data.
{had.co.nz}

\bibitem[\protect\citeauthoryear{Iyer et~al.}{Iyer et~al.}{2011}]{LIGOIndia}
Iyer B.,  et~al., 2011, Technical Report LIGO-M1100296, {{LIGO-India}},
  Proposal of the Consortium for Indian Initiative in Gravitational-Wave
  Observations ({{IndIGO}})

\bibitem[\protect\citeauthoryear{Janquart, Hannuksela, Haris  \& Van
  Den~Broeck}{Janquart et~al.}{2021}]{janquartFastPreciseMethodology2021}
Janquart J.,  Hannuksela O.~A.,  Haris K.,   Van Den~Broeck C.,  2021, arXiv
  e-prints, 2105, arXiv:2105.04536

\bibitem[\protect\citeauthoryear{Kim, Lee, Yuen, Hannuksela  \& Li}{Kim
  et~al.}{2021}]{Kim:2020xkm}
Kim K.,  Lee J.,  Yuen R. S.~H.,  Hannuksela O.~A.,   Li T. G.~F.,  2021,
  \mn@doi [ApJ] {10.3847/1538-4357/ac0143}, 915, 119

\bibitem[\protect\citeauthoryear{Lai, Hannuksela, {Herrera-Mart{\'i}n}, Diego,
  Broadhurst  \& Li}{Lai et~al.}{2018}]{Lai:2018rto}
Lai K.-H.,  Hannuksela O.~A.,  {Herrera-Mart{\'i}n} A.,  Diego J.~M.,
  Broadhurst T.,   Li T. G.~F.,  2018, \mn@doi [Phys. Rev. D]
  {10.1103/PhysRevD.98.083005}, 98, 083005

\bibitem[\protect\citeauthoryear{Laureijs et~al.,}{Laureijs
  et~al.}{2011}]{laureijsEuclidDefinitionStudy2011}
Laureijs R.,  et~al., 2011, arXiv e-prints, 1110, arXiv:1110.3193

\bibitem[\protect\citeauthoryear{Li, Mao, Zhao  \& Lu}{Li
  et~al.}{2018}]{Li:2018prc}
Li S.-S.,  Mao S.,  Zhao Y.,   Lu Y.,  2018, \mn@doi [MNRAS]
  {10.1093/mnras/sty411}, 476, 2220

\bibitem[\protect\citeauthoryear{Li, Lo, Sachdev, Chan, Lin, Li  \&
  Weinstein}{Li et~al.}{2019}]{liFindingDiamondsRough2021}
Li A. K.~Y.,  Lo R. K.~L.,  Sachdev S.,  Chan C.~L.,  Lin E.~T.,  Li T. G.~F.,
   Weinstein A.~J.,  2019, arXiv e-prints, 1904, arXiv:1904.06020

\bibitem[\protect\citeauthoryear{Liao, Ding, Biesiada, Fan  \& Zhu}{Liao
  et~al.}{2018}]{liaoAnomaliesTimeDelays2018}
Liao K.,  Ding X.,  Biesiada M.,  Fan X.-L.,   Zhu Z.-H.,  2018, \mn@doi [ApJ]
  {10.3847/1538-4357/aae30f}, 867, 69

\bibitem[\protect\citeauthoryear{Liu, Hernandez  \& Creighton}{Liu
  et~al.}{2021}]{Liu:2020par}
Liu X.,  Hernandez I.~M.,   Creighton J.,  2021, \mn@doi [ApJ]
  {10.3847/1538-4357/abd7eb}, 908, 97

\bibitem[\protect\citeauthoryear{Lo \& Hernandez}{Lo \&
  Hernandez}{2021}]{Lo:2021nae}
Lo R. K.~L.,  Hernandez I.~M.,  2021, ArXiv210409339 Astro-Ph Physicsgr-Qc

\bibitem[\protect\citeauthoryear{Magee et~al.}{Magee
  et~al.}{2019}]{Magee:2019vmb}
Magee R.,  et~al., 2019, \mn@doi [ApJL] {10.3847/2041-8213/ab20cf}, 878, L17

\bibitem[\protect\citeauthoryear{Mapelli}{Mapelli}{2020}]{mapelliBinaryBlackHole2020}
Mapelli M.,  2020, \mn@doi [Frontiers in Astronomy and Space Sciences]
  {10.3389/fspas.2020.00038}, 7, 38

\bibitem[\protect\citeauthoryear{McIsaac, Keitel, Collett, Harry, Mozzon, Edy
  \& Bacon}{McIsaac et~al.}{2020}]{mcisaacSearchStronglyLensed2020}
McIsaac C.,  Keitel D.,  Collett T.,  Harry I.,  Mozzon S.,  Edy O.,   Bacon
  D.,  2020, \mn@doi [Phys. Rev. D] {10.1103/PhysRevD.102.084031}, 102, 084031

\bibitem[\protect\citeauthoryear{Meena \& Bagla}{Meena \&
  Bagla}{2020}]{Meena:2019ate}
Meena A.~K.,  Bagla J.~S.,  2020, \mn@doi [MNRAS] {10.1093/mnras/stz3509}, 492,
  1127

\bibitem[\protect\citeauthoryear{Metcalf et~al.,}{Metcalf
  et~al.}{2019}]{metcalfStrongGravitationalLens2019}
Metcalf R.~B.,  et~al., 2019, \mn@doi [A\&A] {10.1051/0004-6361/201832797},
  625, A119

\bibitem[\protect\citeauthoryear{Mukherjee, Broadhurst, Diego, Silk  \&
  Smoot}{Mukherjee et~al.}{2021}]{mukherjeeImpactAstrophysicalBinary2021}
Mukherjee S.,  Broadhurst T.,  Diego J.~M.,  Silk J.,   Smoot G.~F.,  2021,
  arXiv e-prints, 2106, arXiv:2106.00392

\bibitem[\protect\citeauthoryear{{Nagam}, {Koopmans}, {Valentijn}, {Kleijn},
  {de Jong}, {Napolitano}, {Li}  \& {Tortora}}{{Nagam}
  et~al.}{2023}]{nagamdenselens}
{Nagam} B.~C.,  {Koopmans} L. V.~E.,  {Valentijn} E.~A.,  {Kleijn} G.~V.,  {de
  Jong} J. T.~A.,  {Napolitano} N.,  {Li} R.,   {Tortora} C.,  2023, \mn@doi
  [Monthly Notices of the Royal Astronomical Society] {10.1093/mnras/stad1623},
  \href {https://ui.adsabs.harvard.edu/abs/2023MNRAS.523.4188N} {523, 4188}

\bibitem[\protect\citeauthoryear{Neal}{Neal}{2001}]{nealAnnealedImportanceSampling2001}
Neal R.~M.,  2001, \mn@doi [Stat. Comput.] {10.1023/A:1008923215028}, 11, 125

\bibitem[\protect\citeauthoryear{Ng, Wong, Broadhurst  \& Li}{Ng
  et~al.}{2018}]{Ng:2017yiu}
Ng K. K.~Y.,  Wong K. W.~K.,  Broadhurst T.,   Li T. G.~F.,  2018, \mn@doi
  [Phys. Rev. D] {10.1103/PhysRevD.97.023012}, 97, 023012

\bibitem[\protect\citeauthoryear{Nitz, Capano, Nielsen, Reyes, White, Brown  \&
  Krishnan}{Nitz et~al.}{2019}]{Nitz:2018imz}
Nitz A.~H.,  Capano C.,  Nielsen A.~B.,  Reyes S.,  White R.,  Brown D.~A.,
  Krishnan B.,  2019, \mn@doi [ApJ] {10.3847/1538-4357/ab0108}, 872, 195

\bibitem[\protect\citeauthoryear{Nitz et~al.,}{Nitz
  et~al.}{2020}]{Nitz:2020oeq}
Nitz A.~H.,  et~al., 2020, \mn@doi [ApJ] {10.3847/1538-4357/ab733f}, 891, 123

\bibitem[\protect\citeauthoryear{Nitz et~al.,}{Nitz
  et~al.}{2021a}]{nitzGwastroPycbcPyCBC2021}
Nitz A.,  et~al., 2021a, Gwastro/Pycbc: {{PyCBC Release}} 1.18.1, Zenodo,
  \mn@doi{10.5281/zenodo.4849433}

\bibitem[\protect\citeauthoryear{Nitz, Capano, Kumar, Wang, Kastha,
  Sch{\"a}fer, Dhurkunde  \& Cabero}{Nitz et~al.}{2021b}]{Nitz:2021uxj}
Nitz A.~H.,  Capano C.~D.,  Kumar S.,  Wang Y.-F.,  Kastha S.,  Sch{\"a}fer M.,
   Dhurkunde R.,   Cabero M.,  2021b, \mn@doi [ApJ] {10.3847/1538-4357/ac1c03},
  922, 76

\bibitem[\protect\citeauthoryear{Oguri}{Oguri}{2018a}]{Oguri:2018muv}
Oguri M.,  2018a, \mn@doi [MNRAS] {10.1093/mnras/sty2145}, 480, 3842

\bibitem[\protect\citeauthoryear{Oguri}{Oguri}{2018b}]{oguriEffectGravitationalLensing2018}
Oguri M.,  2018b, \mn@doi [MNRAS] {10.1093/mnras/sty2145}, 480, 3842

\bibitem[\protect\citeauthoryear{Oguri \& Kawano}{Oguri \&
  Kawano}{2003}]{Oguri:2002ku}
Oguri M.,  Kawano Y.,  2003, \mn@doi [MNRAS]
  {10.1046/j.1365-8711.2003.06290.x}, 338, L25

\bibitem[\protect\citeauthoryear{Pagano, Hannuksela  \& Li}{Pagano
  et~al.}{2020}]{Pagano:2020rwj}
Pagano G.,  Hannuksela O.~A.,   Li T. G.~F.,  2020, \mn@doi [A\&A]
  {10.1051/0004-6361/202038730}, 643, A167

\bibitem[\protect\citeauthoryear{Pang, Lo, Wong, Li  \& Van Den~Broeck}{Pang
  et~al.}{2020}]{Pang:2020pfz}
Pang P. T.~H.,  Lo R. K.~L.,  Wong I. C.~F.,  Li T. G.~F.,   Van Den~Broeck C.,
   2020, \mn@doi [Phys. Rev. D] {10.1103/PhysRevD.101.104055}, 101, 104055

\bibitem[\protect\citeauthoryear{Petrillo et~al.,}{Petrillo
  et~al.}{2019}]{petrilloTestingConvolutionalNeural2019}
Petrillo C.~E.,  et~al., 2019, \mn@doi [MNRAS] {10.1093/mnras/sty2683}, 482,
  807

\bibitem[\protect\citeauthoryear{Petrov et~al.,}{Petrov
  et~al.}{2022}]{Petrov:2021bqm}
Petrov P.,  et~al., 2022, \mn@doi [Astrophys. J.] {10.3847/1538-4357/ac366d},
  924, 54

\bibitem[\protect\citeauthoryear{{Planck Collaboration} et~al.,}{{Planck
  Collaboration} et~al.}{2020}]{planckcollaborationPlanck2018Results2020}
{Planck Collaboration} et~al., 2020, \mn@doi [A\&A]
  {10.1051/0004-6361/201833910}, 641, A6

\bibitem[\protect\citeauthoryear{Robertson, Smith, Massey, Eke, Jauzac,
  Bianconi  \& Ryczanowski}{Robertson et~al.}{2020}]{Robertson:2020mfh}
Robertson A.,  Smith G.~P.,  Massey R.,  Eke V.,  Jauzac M.,  Bianconi M.,
  Ryczanowski D.,  2020, \mn@doi [Mon. Not. R. Astron. Soc.]
  {10.1093/mnras/staa1429}, 495, 3727

\bibitem[\protect\citeauthoryear{Ryczanowski, Smith, Bianconi, McGee,
  Robertson, Massey  \& Jauzac}{Ryczanowski et~al.}{2022}]{Ryczanowski:2022int}
Ryczanowski D.,  Smith G.~P.,  Bianconi M.,  McGee S.,  Robertson A.,  Massey
  R.,   Jauzac M.,  2022, arXiv e-prints

\bibitem[\protect\citeauthoryear{Saleem et~al.,}{Saleem
  et~al.}{2022}]{Saleem:2021iwi}
Saleem M.,  et~al., 2022, \mn@doi [Class. Quantum Grav.]
  {10.1088/1361-6382/ac3b99}, 39, 025004

\bibitem[\protect\citeauthoryear{Santoliquido, Mapelli, Giacobbo, Bouffanais
  \& Artale}{Santoliquido et~al.}{2021}]{santoliquidoCosmicMergerRate2021}
Santoliquido F.,  Mapelli M.,  Giacobbo N.,  Bouffanais Y.,   Artale M.~C.,
  2021, \mn@doi [MNRAS] {10.1093/mnras/stab280}, 502, 4877

\bibitem[\protect\citeauthoryear{Schaefer, Geiger, Kuntzer  \& Kneib}{Schaefer
  et~al.}{2018}]{schaeferDeepConvolutionalNeural2018}
Schaefer C.,  Geiger M.,  Kuntzer T.,   Kneib J.-P.,  2018, \mn@doi [A\&A]
  {10.1051/0004-6361/201731201}, 611, A2

\bibitem[\protect\citeauthoryear{Sereno, Jetzer, Sesana  \& Volonteri}{Sereno
  et~al.}{2011}]{Sereno:2011ty}
Sereno M.,  Jetzer P.,  Sesana A.,   Volonteri M.,  2011, \mn@doi [MNRAS]
  {10.1111/j.1365-2966.2011.18895.x}, 415, 2773

\bibitem[\protect\citeauthoryear{Seto}{Seto}{2004}]{Seto:2003iw}
Seto N.,  2004, \mn@doi [Phys. Rev. D] {10.1103/PhysRevD.69.022002}, 69, 022002

\bibitem[\protect\citeauthoryear{Shu et~al.,}{Shu
  et~al.}{2015}]{shuSloanLensACS2015}
Shu Y.,  et~al., 2015, \mn@doi [ApJ] {10.1088/0004-637X/803/2/71}, 803, 71

\bibitem[\protect\citeauthoryear{Singh, Li, Hannuksela, Li  \& Kim}{Singh
  et~al.}{2019}]{Singh:2018csp}
Singh A.~J.,  Li I. S.~C.,  Hannuksela O.~A.,  Li T. G.~F.,   Kim K.,  2019,
  \mn@doi [AJUR] {10.33697/ajur.2019.019}, 16, 5

\bibitem[\protect\citeauthoryear{Smith, Jauzac, Veitch, Farr, Massey  \&
  Richard}{Smith et~al.}{2018}]{Smith:2017mqu}
Smith G.~P.,  Jauzac M.,  Veitch J.,  Farr W.~M.,  Massey R.,   Richard J.,
  2018, \mn@doi [MNRAS] {10.1093/mnras/sty031}, 475, 3823

\bibitem[\protect\citeauthoryear{Smith, Robertson, Bianconi  \& Jauzac}{Smith
  et~al.}{2019a}]{Smith:2019dis}
Smith G.~P.,  Robertson A.,  Bianconi M.,   Jauzac M.,  2019a, ArXiv190205140
  Astro-Ph

\bibitem[\protect\citeauthoryear{{Smith} et~al.,}{{Smith}
  et~al.}{2019b}]{Smith:2019polarisation}
{Smith} G.~P.,  et~al., 2019b, \mn@doi [Monthly Notices of the Royal
  Astronomical Society] {10.1093/mnras/stz675}, \href
  {https://ui.adsabs.harvard.edu/abs/2019MNRAS.485.5180S} {485, 5180}

\bibitem[\protect\citeauthoryear{Smith et~al.,}{Smith
  et~al.}{2022}]{Smith:2022vbp}
Smith G.~P.,  et~al., 2022, arXiv e-prints

\bibitem[\protect\citeauthoryear{Somiya}{Somiya}{2012}]{Somiya:2011np}
Somiya K.,  2012, \mn@doi [Class. Quant. Grav.]
  {10.1088/0264-9381/29/12/124007}, 29, 124007

\bibitem[\protect\citeauthoryear{Sygnet, Tu, Fort  \& Gavazzi}{Sygnet
  et~al.}{2010}]{sygnetSearchEdgeonGalaxy2010}
Sygnet J.~F.,  Tu H.,  Fort B.,   Gavazzi R.,  2010, \mn@doi [A\&A]
  {10.1051/0004-6361/200913977}, 517, A25

\bibitem[\protect\citeauthoryear{Tessore \& Benton~Metcalf}{Tessore \&
  Benton~Metcalf}{2015}]{tessoreEllipticalPowerLaw2015}
Tessore N.,  Benton~Metcalf R.,  2015, \mn@doi [A\&A]
  {10.1051/0004-6361/201526773}, 580, A79

\bibitem[\protect\citeauthoryear{Tessore \& Metcalf}{Tessore \&
  Metcalf}{2016}]{tessoreEllipticalPowerLaw2016}
Tessore N.,  Metcalf R.~B.,  2016, \mn@doi [A\&A]
  {10.1051/0004-6361/201526773e}, 593, C2

\bibitem[\protect\citeauthoryear{{The LIGO Scientific Collaboration}
  et~al.,}{{The LIGO Scientific Collaboration}
  et~al.}{2021}]{LIGOScientific:2021djp}
{The LIGO Scientific Collaboration} et~al., 2021, arXiv e-prints, \href
  {https://ui.adsabs.harvard.edu/abs/2021arXiv211103606T} {p. arXiv:2111.03606}

\bibitem[\protect\citeauthoryear{Torrey et~al.,}{Torrey
  et~al.}{2015}]{torreyAnalysisEvolvingComoving2015}
Torrey P.,  et~al., 2015, \mn@doi [MNRAS] {10.1093/mnras/stv1986}, 454, 2770

\bibitem[\protect\citeauthoryear{Venumadhav, Zackay, Roulet, Dai  \&
  Zaldarriaga}{Venumadhav et~al.}{2019}]{Venumadhav:2019tad}
Venumadhav T.,  Zackay B.,  Roulet J.,  Dai L.,   Zaldarriaga M.,  2019,
  \mn@doi [Phys. Rev. D] {10.1103/PhysRevD.100.023011}, 100, 023011

\bibitem[\protect\citeauthoryear{Venumadhav, Zackay, Roulet, Dai  \&
  Zaldarriaga}{Venumadhav et~al.}{2020}]{Venumadhav:2019lyq}
Venumadhav T.,  Zackay B.,  Roulet J.,  Dai L.,   Zaldarriaga M.,  2020,
  \mn@doi [Phys. Rev. D] {10.1103/PhysRevD.101.083030}, 101, 083030

\bibitem[\protect\citeauthoryear{Vogelsberger et~al.,}{Vogelsberger
  et~al.}{2014}]{vogelsbergerPropertiesGalaxiesReproduced2014}
Vogelsberger M.,  et~al., 2014, \mn@doi [Nature] {10.1038/nature13316}, 509,
  177

\bibitem[\protect\citeauthoryear{Wierda, Wempe, Hannuksela, Koopmans  \& Van
  Den~Broeck}{Wierda et~al.}{2021}]{wierdaDetectorHorizonForecasting2021}
Wierda A. R. A.~C.,  Wempe E.,  Hannuksela O.~A.,  Koopmans L. V.~E.,   Van
  Den~Broeck C.,  2021, arXiv e-prints, 2106, arXiv:2106.06303

\bibitem[\protect\citeauthoryear{Williams et~al.,}{Williams
  et~al.}{2018}]{williamsJWSTExtragalacticMock2018}
Williams C.~C.,  et~al., 2018, \mn@doi [ApJS] {10.3847/1538-4365/aabcbb}, 236,
  33

\bibitem[\protect\citeauthoryear{Xu, Ezquiaga  \& Holz}{Xu
  et~al.}{2021}]{Xu:2021bfn}
Xu F.,  Ezquiaga J.~M.,   Holz D.~E.,  2021, ArXiv210514390 Astro-Ph

\bibitem[\protect\citeauthoryear{Yang, Wu, Liao, Ding, You, Cao, Biesiada  \&
  Zhu}{Yang et~al.}{2021}]{yangEventRatePredictions2021}
Yang L.,  Wu S.,  Liao K.,  Ding X.,  You Z.,  Cao Z.,  Biesiada M.,   Zhu
  Z.-H.,  2021, arXiv e-prints, 2105, arXiv:2105.07011

\bibitem[\protect\citeauthoryear{Yu, Zhang  \& Wang}{Yu
  et~al.}{2020}]{Yu:2020agu}
Yu H.,  Zhang P.,   Wang F.-Y.,  2020, \mn@doi [MNRAS]
  {10.1093/mnras/staa1952}, 497, 204

\bibitem[\protect\citeauthoryear{Zackay, Dai, Venumadhav, Roulet  \&
  Zaldarriaga}{Zackay et~al.}{2019a}]{Zackay:2019btq}
Zackay B.,  Dai L.,  Venumadhav T.,  Roulet J.,   Zaldarriaga M.,  2019a,
  \mn@doi [ArXiv191009528 Astro-Ph Physicsgr-Qc] {10.1103/PhysRevD.104.063030}

\bibitem[\protect\citeauthoryear{Zackay, Venumadhav, Dai, Roulet  \&
  Zaldarriaga}{Zackay et~al.}{2019b}]{Zackay:2019tzo}
Zackay B.,  Venumadhav T.,  Dai L.,  Roulet J.,   Zaldarriaga M.,  2019b,
  \mn@doi [Phys. Rev. D] {10.1103/PhysRevD.100.023007}, 100, 023007

\makeatother
\end{thebibliography}

\appendix

\section{Realistic Euclid images}
\label{sec:euclidsnr}
According to \citet{cropperVISVisibleImager2016}, for a VIS magnitude of 25.0, a point source is expected to reach a signal to noise of $\text{SNR} = 10$ when doing aperture photometry with an aperture diameter of \SI{1.3}{\arcsec} (quite a bit better than the requirements).
The aperture photometry signal-to-noise ratio (SNR) is given by
\begin{align}
    \text{SNR} = \sqrt{N_\text{exp}}\frac{F_\text{ap}}{\sqrt{N_\text{pix}\sigma^2}},
\end{align}
where $N_\text{exp}$ is the number of exposures, we assume $N_\text{exp}=3$.

For the rest, we follow mostly the approach of \citet{euclidcollaborationEuclidPreparationIV2019}, with some minor differences.
\begin{align}
F_\text{ap} &= t_\text{exp} 10^{-(m-\text{ZP})/2.5} \\
\sigma^2 &= F_\text{sky} + \sigma_\text{ro}^2\\
F_\text{sky} &= t_\text{exp} \delta^2 10^{-(\mu_\text{sky}-\text{ZP})/2.5},
\end{align}
where \begin{description}
    \item $m$ is the object's magnitude (25.0 for SNR = 10)
    \item ZP is the photometric zero-point (conversion from magnitude to electron counts)
    \item $\sigma_\text{ro} = \SI{4.2}{e^-/pix}$ is the readout noise
    \item $\delta = \SI{0.1}{\arcsec}$ is the pixel scale
    \item $t_\text{exp} = \SI{565}{s}$ is the exposure time
    \item $\mu_\text{sky} = \SI{22.35}{mag/arcsec^2}$ is the sky brightness
\end{description}
The aperture correction due to not getting all the signal in the \SI{1.3}{\arcsec} aperture negligible due to the small FWHM, $f = \exp(-\frac{(1.3/2)^2}{2\cdot 0.17^2/8\log{2}}) = \num{2e-18}$.
Setting $\text{SNR}=10$ at $m=25$, gives a photometric zeropoint of $ZP=25.12$ and a background noise of $\sigma_\text{bg}=\sqrt{F_\text{sky}+\sigma_\text{ro}^2} = \SI{9.494}{e^-/pix}$ for each image. The 3-image coadd would therefore have $\sigma_\text{bg} = \sqrt{3}\cdot\SI{9.494}{e^-/pix}$, with a total exposure time of $3 t_\text{exp}=\SI{1695}{s}$ (due to the dithering pattern, half of the area is in fact 4 exposures, but this is neglected; see \citet{collettPopulationGalaxyGalaxyStrong2015}). The same calculation by \citet{euclidcollaborationEuclidPreparationIV2019} yields $ZP=24.0$, with the difference arising because (i) the authors using the design sensitivity instead of the predicted sensitivity, (ii) the authors considering the calculation as a single exposure of \SI{1695}{s} instead of 3 exposures of \SI{565}{s} and (iii) the authors assume that the SNR of 10 is achieved with good SExtractor settings while in the \citet{euclidcollaborationEuclidPreparationIV2019} simulations, it was actually achieved with very sub-optimal aperture photometry. \citet{collettPopulationGalaxyGalaxyStrong2015} uses $ZP=25.5$. We note that the results are robust against reasonable variation of the assumed \textit{Euclid} noise parameters.

\section{Observability by Euclid}
\label{app:observability}
\begin{figure*}
    \centering
    \includegraphics[width=\linewidth]{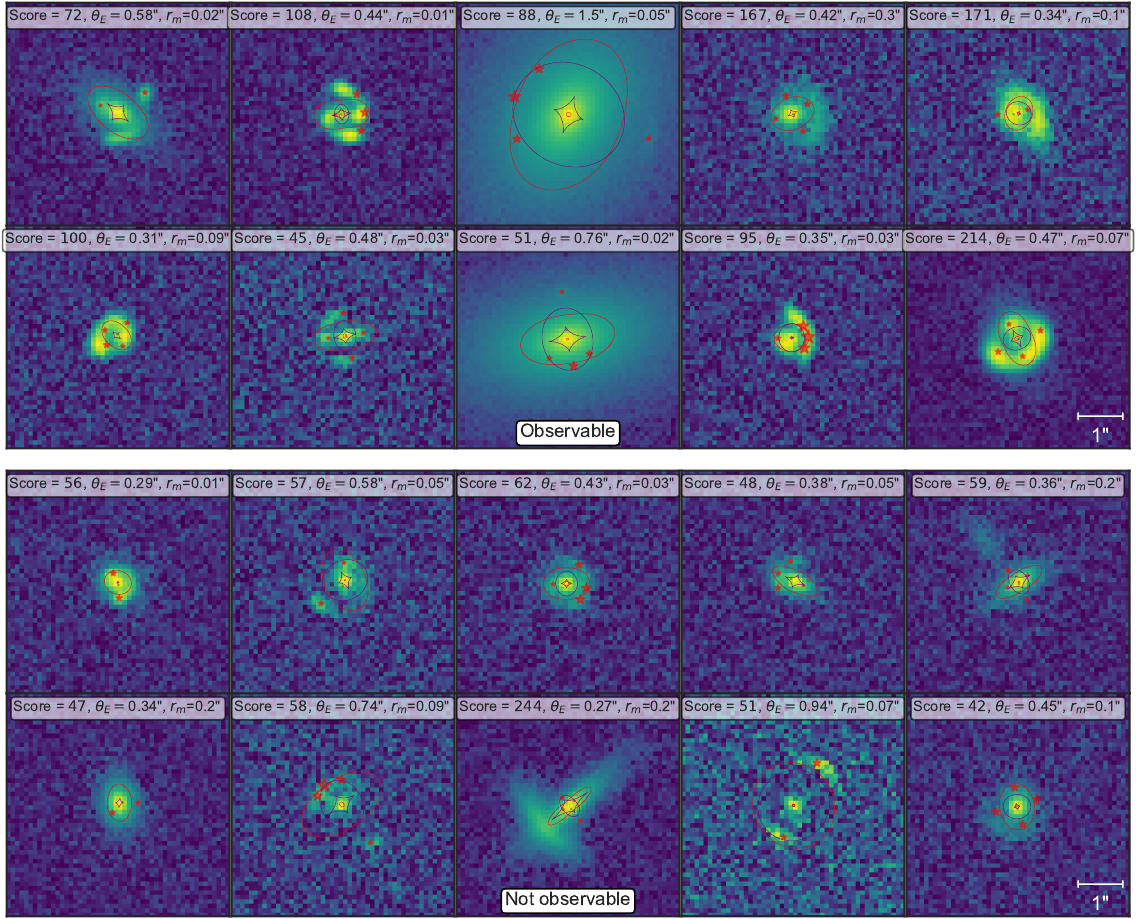}
    \caption{Some example classifications on lens images out of the visually classified sample, separated by detectable lenses (top 10) and non-detectable lenses (bottom 10). Extra lines and markers as in \cref{fig:panel}.}
    \label{fig:exampleclassifications}
\end{figure*}
\begin{figure}
    \centering
    \includegraphics{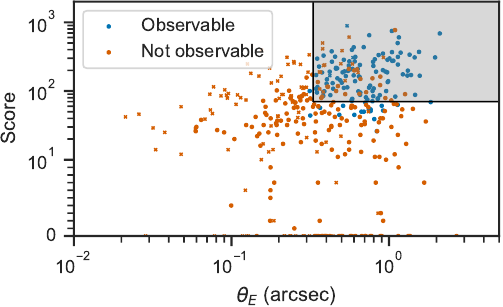}
    \caption{Classification thresholds. Each galaxy is a blue or orange point, depending on whether it was deemed observable. If a galaxy did not pass the size threshold (semi-minor axis length $r_m>0.56\theta_E$), the dot was replaced with a cross.}
    \label{fig:classificationthresholds}
\end{figure}
To obtain realistic requirements for classifying an image as observable (that is, detectable and recognisable as a lens by a good lens finder, \cref{sec:imgreq}), a visual inspection of a subset of the simulated systems was done. A human classification was carried out for several hundred systems to assess whether it would be observable. Some example classifications (in particular some of the harder ones) are shown in \cref{fig:exampleclassifications}, where the top panel images were classified as lenses that have clear lensing features that a lens finder may pick up, whereas the images in the bottom panel are assessed as too bad for identification. Since some level of subjectivity is involved, we tried to remain conservative but are helped by the fact that \textit{Euclid} has both a high spatial resolution and a relatively high SNR for most potential lenses that could lens GW events. The detectability seems to rely mainly on the Einstein radius and the background source information content (which is summarised as a score: the number of lensed-image pixels with $\text{SNR}>1.5$, see \cref{sec:imgreq}). The relation between these parameters and observability is shown in \cref{fig:classificationthresholds}. It can be seen that a reasonable cut can be made by setting hard thresholds on these parameters, indicated with the gray region. As can be seen in a few of the lenses in the lower panel of \cref{fig:exampleclassifications}, there is a group of lens systems that is not observable because the background galaxy is too smooth, and therefore, lensing has too little visible impact on the image morphology. Because of this, we include a cut on the source galaxy semi-minor axis $r_m$, in \cref{fig:exampleclassifications}, galaxies that did not pass this threshold are indicated with crosses. The final thresholds ($\theta_E>\SI{0.33}{\arcsec}$, $\text{Score}>70$ and $r_m<0.56\theta_E$) were found by maximising the accuracy with respect to the human classification while keeping the total fraction of detectable lenses in the visually inspected sample the same. More sophisticated lens finders are available, but this is left for future work.

\section{Magnification distributions}
\label{app:magnification}
\begin{figure}
    \centering
    \includegraphics[width=\linewidth]{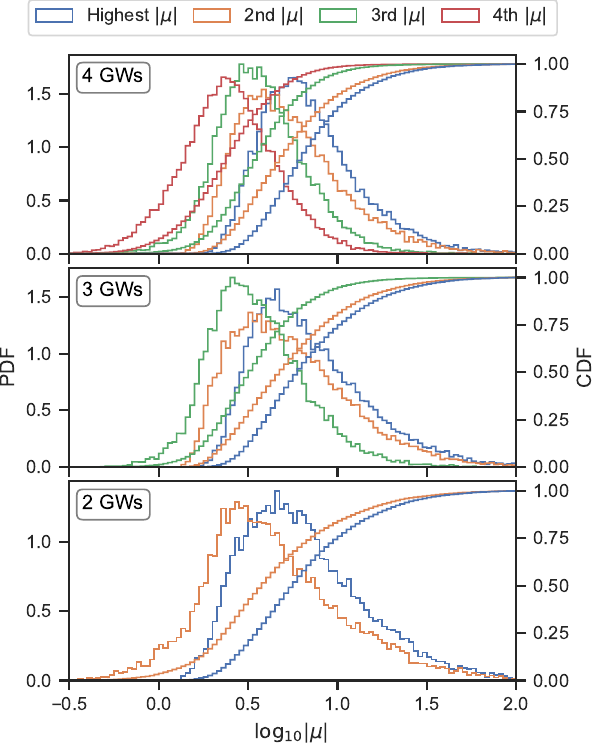}
    \caption{Distributions of the absolute value of the magnification of the different GW signals. The top panel shows magnification distributions for the GWs in case 4 GWs were detected, the middle panel when 3 GWs were detected and the lower panel when 2 GWs were detected. The different colors indicate the different images, ordered from highest magnification (blue) to second highest (yellow), third highest (green), and fourth highest/lowest (red). The left axes show the probability density (PDF), and the right axes the cumulative distribution (CDF).}
    \label{fig:magnificationdistr}
\end{figure}

Due to the magnification bias, higher magnification systems are over-represented compared to the astrophysical distribution in samples of lenses that are SNR-limited (or similar). It is therefore interesting to quantify how strong this effect is for our lensed gravitational waves, for instance, if the sample is dominated by high-magnification systems, or more average lens systems. For our simulations done in \cref{sec:samplegen}, we have put observability limits on the GW signals. The distributions, for the systems conditional on the GWs being detected (\cref{eq:plensgwobs}), are shown in \cref{fig:magnificationdistr}. 
We note that in systems of high magnification, only 2 out of 4 images are usually highly magnified, and that because of this, when requiring 3 or 4 GWs to be detected, the magnification bias is not very strong. For cases where 2 GWs are detected, there is a more substantial high-magnification tail, but in total still only \SI{\sim20}{\percent} of systems have magnifications above \num{10}.

\section{Ground-based follow-up}\label{sec:groundbasedfollowup}
In case the gravitational wave is outside of \textit{Euclid} coverage, a ground-based follow-up may still be possible. To investigate this, we run the simulations and the identification tests also for imaging with a ground-based telescope. Specifically, we test our framework with Subaru/HSC imaging (although different telescopes, for instance a target-of-opportunity follow-up with Rubin/LSST on a night with good seeing would be possible too.). We (conservatively) assume the imaging depth parameters achieved in the r-band by the `Ultra-deep survey' of the Subaru Hiper-Suprime Cam Strategic Program \citep{aiharaFirstDataRelease2018}. Since seeing is the most important parameter, we consider two cases, one with good seeing of \SI{0.4}{\arcsec} and one where the seeing is \SI{0.6}{\arcsec}. The resulting accuracies are shown in \cref{tab:rates_hsc}.
Although the numbers are quite a bit lower than in the Euclid-based analysis, in this ground-based follow-up scenario, we have only used single-band information. 
It is likely that multi-band information helps, especially on the lens finding because one can make use of the difference in colour of the lens and source galaxies \citep{collettPopulationGalaxyGalaxyStrong2015}.

\begin{table*}
    \centering
    \caption{Similar to \cref{tab:rates}, but for a ground-based follow-up. We consider only $p(\text{GW emitted})\propto\text{SFR}$, and an all-galaxy velocity dispersion function. Numerical uncertainties on the lensing counts are about \SI{10}{\percent}, but the accuracies suffer from low-number statistics.
    }
    \label{tab:rates_hsc}
\begin{tabular}{
  l
  r
  S[table-format=1.2e1]
  S[table-format=1.2]
  *{1}{%
   S[table-format=2.1]
   S[table-format=2.1]
   S[table-format=2.1]
  }
}
\toprule
{Seeing} & {Images} &  {\splitcell{EM lenses}} & \multicolumn{4}{c}{lensed GWs, SFR} \\
& & & \si{yr^{-1}} & {\si{\percent} obs.} & {\si{\percent} acc.} & {\si{\percent} $>2\sigma$}\\
\midrule
\multirow{3}{*}{$s=\SI{0.4}{\arcsec}$} 
&\num{\ge2} & 9.92e5 & 5.05 & 13.9 & 0.1 & 0.0\\
&\num{\ge3} & 1.89e5 & 1.15 & 10.9 & 4.0 & 1.3\\
&\num{\ge4} & 1.83e5 & 0.43 & 8.9 & 1.3 & 0.5\\
\midrule
\multirow{3}{*}{$s=\SI{0.6}{\arcsec}$}
&\num{\ge2} & 3.5e5 & 5.05 & 7.6 & 3.0 & 1.1\\
&\num{\ge3} & 9.4e4 & 1.15 & 5.7 & 6.7 & 5.2\\
&\num{\ge4} & 7.3e4 & 0.43 & 4.6 & 1.9 & 0.7\\
\bottomrule
\end{tabular}
\end{table*}

\section{Pairing GWs to lensed images}
\label{app:pairingimages}
A priori, the pairings between model time-delays and flux ratios and measured time delays and flux ratios are not known. Unlike in quasar or supernova time-delay cosmography, there is no information that links a measured time delay time to a specific pair of images. An apparent fix would be to just sort the images based on their arrival times. However this ordering is not necessarily correct, or unique, because of the (systematic) uncertainties on the arrival times. In some lens configurations, this is problematic, because the sorting of measured arrival times leads to a wrong time-to-image correspondence. Ideally, one would add another parameter to the model: the permutation $p$ that links each image to an arrival time and magnification. Since we are not directly interested in it and sampling this discrete parameter explicitly is impractical, this permutation is marginalised over. This gives
\begin{equation}
    \mathcal{L}(\vb*y_\text{GW}|\vb*\theta) = \sum_{\text{permutation }p}\mathcal{L}(\vb* y_\text{GW}|\vb*\theta,p).
\end{equation}
The information linking the BBH position distribution to the light distribution and the assumption that the probability of a GW is proportional to the luminosity is already incorporated in the priors.

\section{Constructing the time-delay likelihood}
\label{app:margdelay}
Since the absolute start time of a set of events has a flat prior, it can be marginalised over. Assuming that the right events can be coupled to the right images, let $\vb*{t}_\text{meas}$ be the measured times. Let $\vb*{t}_\text{mod}$ be the model times. Then the residuals are $\vb*{x} = \vb*{t}_\text{meas} - \vb*{t}_\text{mod}$. Assume Gaussian errors on the time-delays. Because the absolute time needs to be subtracted from these residuals, the likelihood is:
\begin{align}
    \mathcal{L}(\text{time delays}) = \int_{-\infty}^\infty\dd{t_0}\det(2\uppi\vb*\Sigma)^{-1/2}\exp[-\frac{1}{2}(\vb*x-t_0)^T\vb*\Sigma^{-1}(\vb*x-t_0)],
\end{align}
where $t_0$ is a time offset. With a diagonal covariance matrix, this can be worked out further. Let $u=t_0-x_0$, and $\vb*x'[n] = x[1...(N-1)]-x[0]$ where $x_0$ is the time-delay residual of the first image, and $N$ is the number of images.
\begin{align}
    \mathcal{L}&= \int_{-\infty}^\infty\dd{u}(2\uppi\sigma^2)^{-N/2}\exp[\frac{-1}{2\sigma^2}(\vb*x-x_0-u)^T(\vb*x-x_0-u)]\\
    &=(2\uppi\sigma^2)^{-N/2}\int_{-\infty}^\infty\dd{u}\exp[\frac{-1}{2\sigma^2}\qty(\mqty(0\\\vb*x')-u)^T\qty(\mqty(0\\\vb*x')-u)]\\
    &=(2\uppi\sigma^2)^{-N/2}\exp[-\frac{\vb*x'\cdot\vb*x'}{2\sigma^2}]\int_{-\infty}^\infty\dd{u}\exp[\frac{-1}{2\sigma^2}\qty(-2(\sum_ix'_i)u+u^2N)]
\end{align}
Using the standard integral $\int_{-\infty}^\infty\dd{x}e^{-(ax^2+bx)} = \sqrt{\frac{\uppi}{a}}e^{\frac{b^2}{4a}}$ this can be worked out further:
\begin{align}
    \mathcal{L}&=(2\uppi\sigma^2)^{-N/2}\sqrt{\frac{\uppi\cdot 2\sigma^2}{N}}\exp[-\frac{\vb*x'\cdot\vb*x'}{2\sigma^2}]\exp[\frac{(\sum_ix'_i)^2}{N\cdot 2\sigma^2}]\\
    &=(2\uppi\sigma^2)^{-(N-1)/2}\sqrt{\frac{1}{N}}\exp[-\frac{1}{2\sigma^2}({x'_i(\delta_{ij}-\frac{1}{N})x'_j})]
\end{align}
This is a Normal distribution with inverse covariance matrix $\vb*\Sigma^{-1}=\frac{1}{\sigma^2}(\vb*I_{N-1}-\frac{1}{N})$, where $\vb*I_n$ is the $n\times n$ identity matrix. Inverting this (using the Sherman-Morrison formula) yields $\vb*\Sigma=\sigma^2(\vb*I_{N-1}+1)$, or $(\vb*\Sigma)_{ij} =\sigma^2+\sigma^2\delta_{ij}$. The determinant can be calculated with Sylvester's theorem, which yields $\det(\vb*\Sigma)=(\sigma^2)^{N-1}N$, confirming that the derived distribution is normalised, as it should be.

\section{A fast caustics calculation}
\label{sec:caustics}
\citet{tessoreEllipticalPowerLaw2015} do not give a fast (analytical) way of calculating the critical curves (and therefore the caustics). \texttt{lenstronomy} does include several methods of computing the caustics but its numerical uncertainty makes the resulting polygon inconvenient to sample from (for example, there will be unphysical self-intersections). A small description of a reliable analytical procedure for calculating critical curves for an elliptical power law + shear model is therefore shortly described here and has since been added to \texttt{lenstronomy}. The derivation involves adding an extra external shear to equation (17) of \citet{tessoreEllipticalPowerLaw2016} and calculating the inverse magnification. This results in a quadratic equation in $(b/R)^t$, depending on the polar angle $\theta$.

\begin{align*}
    &\mu^{-1} = 1-\abs{\gamma}^2 \\&+ \qty(\frac{b}{R})^t \qty(-(2-t) - 2 \langle \gamma, {-e^{2i\theta}\frac{2-t}{2}+(1-t)e^{i\theta}\frac{2}{1+q}\Omega(\theta) f^{-1}}\rangle)  \\&+ \qty(\frac{b}{R})^{2t}\qty((1-t)(2-t){\langle e^{i\theta}, \Omega(\theta)\rangle}f^{-1}\frac{2}{1+q} - \qty(\frac{2(1-t)}{1+q})^2\abs{\Omega(\theta)}^2f^{-2}),
\end{align*}
where the notation of \cite{tessoreEllipticalPowerLaw2015} was followed, with the additions
\begin{description}
    \item $\langle x, y \rangle = \Re(x)\Re(y)+\Im(x)\Im(y)$
    \item $\gamma=\gamma_1+i\gamma_2$ is the external shear, centred at the ellipse centroid.
    \item $f=r/R$ is the ratio between the polar and elliptical radius.
\end{description}

To calculate the critical curve, a grid of $\theta$ is sampled, and for each $\theta$, the elliptical radius $R$ is solved for. These are combined to get the polar coordinate pairs.
The algorithm was also contributed to lenstronomy, significantly outperforming \texttt{lenstronomy}'s existing critical curve calculations in both speed and precision.

\section{A fast lens equation solver}
\label{sec:lensequationsolving}

In our methodology there is a need to solve lens equation quickly and reliably. The lens equation is 
\begin{equation}
    \vb*{y} = \vb*{x} + \vb*{\alpha}(\vb*{x}),
\end{equation} where $\vb*x$ is the source position vector in the lens plane, $\vb*y$ is the source position vector in the source plane, and $\vb*{\alpha}$ is the deflection. The default existing method in \texttt{lenstronomy} works by first ray-shooting a grid of $\vb*{x}$, finding the local minima on this grid, and refining the candidate solutions. However, close to the caustics this often leads to solutions that are missed unless the grid is very fine, which is computationally expensive. Since images close to caustics are high-magnification, the lens equation solver's being inaccurate for these systems might be especially problematic.

We developed an improved method specialised for the elliptical power law plus shear model which is summarised here. We use that one can solve for radial coordinate $R$ analytically given a polar angle $\phi$. Specifically, we have:
\begin{equation}
    \vb*{y} = \vb*{x} + \vb*{\Gamma}\vb*{x} + \frac{2b}{1+q}\qty(\frac{b}{R})^{t-1} \vb*{\Omega}(\theta), \label{eq:lenseq}
\end{equation}
with shear matrix $\vb*\Gamma = \mqty(\gamma_1 & \gamma_2\\\gamma_2&\gamma_1)$. We now consider the component of this equation orthogonal to the vector $(1+\vb*{\Gamma})\vb*{x}$, i.e. we multiply both sides with this vector rotated by 90 degrees, $\vb*{v}_\perp = \vb{R}(1+\vb*{\Gamma})\vb*{x}$, then the $\vb*x$ term drops, and we can solve for $R$:
\begin{equation}
    \langle \vb*y, \vb*v_\perp \rangle = \frac{2b}{1+q}\qty(\frac{b}{R})^{t-1} \langle \vb*\Omega, \vb*v_\perp \rangle
\end{equation}
For the case where $t$ is close to $1$, we run into a $1/(1-t)$ term in the exponent, so then we consider the component orthogonal to $\vb*{\Omega}$ instead.

To fully solve the lens equation, we must now find $\theta$ such that also the component parallel to the $(1+\vb*{\Gamma})\vb*{x}$ vector in \cref{eq:lenseq} is satisfied. We are careful to avoid any singularities. 
To find the roots of $\theta$, we sample 200 linearly spaced points, and at the locations where the sign of our objective function changes, we do a bracketed search with scipy's \texttt{brentq} to find the precise solution. We furthermore refine our search around the shear-only solution $(1+\vb*{\Gamma})^{-1}\vb*{y}$, and at the extrema of the objective function (where the derivative changes sign) to ensure we do not miss any solutions.

This recipe was tested to be more reliable (especially near caustics) and an order of magnitude faster than the existing routines in \texttt{lenstronomy} for an elliptical power law with external shear. We have since contributed this method to \texttt{lenstronomy}. 

\bsp	%
\label{lastpage}
\end{document}